\documentclass[12pt]{article}
\usepackage{amssymb}
\def\inh{\vskip 0.075truein \noindent\hangindent=12 pt \hangafter=1}

\usepackage{amsthm}\theoremstyle{remark}

\newcommand{\bte}{\begin{quote}\begin{theorem}}
\newcommand{\ete}[1]{\label{#1}\end{theorem}\end{quote}}
\newcommand{\bcom}{\begin{quote}\begin{comment}}
\newcommand{\ecom}[1]{\label{#1}\end{comment}\end{quote}}
\newcommand{\bex}{\begin{quote}\begin{example}}
\newcommand{\eex}[1]{\label{#1}\end{example}\end{quote}}
\newcommand{\bcon}{\begin{quote}\begin{conclusion}}
\newcommand{\econ}[1]{\label{#1}\end{conclusion}\end{quote}}
\newcommand{\bdefi}{\begin{quote}\begin{definition}}
\newcommand{\edefi}[1]{\label{#1}\end{definition}\end{quote}}

\newcommand{\blem}{\begin{quote}\begin{lemma}}
\newcommand{\elem}[1]{\label{#1}\end{lemma}\end{quote}}

\newcommand{\bpr}{\begin{quote}\begin{problem}}
\newcommand{\epr}[1]{\label{#1}\end{problem}\end{quote}}

\newcommand{\f}{\frac}

\newcommand{\n}{\nonumber \\}
\newcommand{\inti}{\int_{-\infty}^\infty}
\newcommand{\beq}{\begin{eqnarray}}
\newcommand{\eeq}[1]{\label{#1}\end{eqnarray}}
\newcommand\eq[1]{(\ref{#1})}
\newcommand{\bfi}{\begin{figure}[24]}
\newcommand{\efi}[1]{\caption{\label{#1}}\end{figure}}
\newcommand\fig[1]{Fig.~\ref{#1}}
\newcommand{\res}{respectively}
\newcommand\gl{\left}
\newcommand\gr{\right}
\newcommand{\CE}{{\cal E}}

\newcommand{\CG}{{\cal G}}

\newcommand{\CR}{{\cal R}}

\newcommand{\CU}{{\cal U}}

\newcommand{\Ga}{\alpha}

\newcommand{\Gd}{\delta}
\newcommand{\Ge}{\varepsilon}

\newcommand{\Gg}{\gamma}

\newcommand{\Gk}{\varkappa}

\newcommand{\Gn}{\eta}
\newcommand{\Gm}{\mu}

\newcommand{\Gj}{\tau}

\newcommand{\GD}{\Delta}

\newcommand{\az}[1]{Sect.$\!$ \ref{#1}}
\newcommand\D{\,\mathrm{d}}
\newcommand\I{\mathrm{i}}
\newcommand\E{\mathrm{e}}
\newcommand{\bexe}{\begin{quote}\begin{exercise}\inh}
\newcommand{\eexe}[1]{\label{#1}\end{exercise}\end{quote}}

\usepackage{graphics,graphicx}
\usepackage{color}
\begin{document}
{\large
\title{Brittle fracture in a periodic structure with internal potential energy. Spontaneous crack propagation.
}}

\author{Mark Ayzenberg-Stepanenko$^{a}$, Gennady Mishuris$^{b}$, Leonid Slepyan$^{c,b}$,}
\date{\small{$^a${\em Department of Mathematics, Ben Gurion University, Beer-Sheva 84105 Israel} \\
 $^b${\em
Institute of Mathematics and Physics,
Aberystwyth University\\
Ceredigion SY23 3BZ
Wales UK} \\
$^c${\em School of Mechanical Engineering, Tel Aviv University\\
P.O. Box 39040, Ramat Aviv 69978 Tel Aviv, Israel
}
}}

\maketitle

\vspace{5mm}\noindent
{\bf Abstract}
Spontaneous brittle fracture is studied based on the recently introduced model (Mishuris and Slepyan, Brittle fracture in a periodic structure with internal potential energy. Proc. Roy. Soc. A, in press). A periodic structure is considered, where only the prospective crack-path layer is specified as a discrete set of alternating initially stretched and compressed bonds. A bridged crack destroying initially stretched bonds may propagate under a certain level of the internal energy without external sources. The general analytical solution with the crack speed $-$ energy relation is presented in terms of the crack-related dynamic Green's function. For the anisotropic two-line chain and lattice considered earlier in quasi-statics, the dynamic problem is examined in detail. The crack speed is found to grow unboundedly as the energy approaches its upper limit. It is revealed that the spontaneous fracture can occur in the form of a pure bridged, partially bridged or fully open crack depending on the internal energy level. Generally, the steady-state mode of the crack propagation is found to be realised, whereas an irregular growth, clustering and the crack speed oscillations are detected in a vicinity of the lower bound of the energy.

\section{Introduction}
We consider a spontaneous crack, propagating in a structural elastic body with periodically distributed, self-equilibrated, microlevel stresses. No external forces are assumed to be applied. The analysis is based on the model introduced in Mishuris and Slepyan (2014), where the static states were examined. In the general formulation, an unspecified periodic structure with a layer of symmetry is considered. Only the latter is specified as a discrete set of alternating stretched and compressed bonds. In particular, such an incompatible stress distribution may occur if the bonds are of different initial lengths.  A general analytical solution corresponds to the bridged crack propagating with a constant speed and destroying the initially stretched bonds. The solution is obtained in terms of the crack-related dynamic Green's function for an unspecified periodic structure. The selective discrete Fourier transform introduced in Mishuris and Slepyan (2014) is used here in dynamics.

The solution represents the internal energy level as a uniquely defined function of the crack speed. The inverse, multi-valued function usually is uniquely defined by the admissibility condition, which states, in general, that only the maximal speed corresponding to the given energy is realized (Marder and Gross, 1995). However, if two or more close minima of the energy exist corresponding to different values of the speed, the latter can oscillate. The revealed phenomenon of instability manifests itself in the case of a two-line chain considered below.

Along with the general structure, two specified structures, the mass-spring lattice and two-line chain are considered in detail. For these structures quantitative results are obtained based on the general solution and the specific Green's functions. We also present results of the numerical simulations of the corresponding transient problems. This allows us to determine the regions of stability of the analytical solutions and to reveal the unsteady regimes as well as partially bridged and fully open crack propagation modes. For these structures the analytical results are specified for an arbitrary value of the parameter defining the structures' anisotropy. Note that the anisotropy plays a substantial role in this problem.

There are two bounds of the internal energy, and the spontaneous crack can exist in the energy segment between them. It cannot propagate if the energy is below the lower bound, whereas the intact structure cannot exist if the energy exceeds the upper bound. The corresponding speed - energy relations are obtained analytically and plotted for the specified structures. The speed range extends from a nonzero value and, in the considered model, has no upper bound. The speed tends to infinity as the energy level approaches its upper bound. The numerical simulations show that, at a high level of the internal energy, not only the stretched bonds but also the initially compressed will break. So, the spontaneous failure wave can propagate in the form of a pure bridged, partially bridged or fully open crack depending on the internal energy level.

Note that in some respects, the problem is related to that for the bridge crack (Mishuris et al, 2008, 2009), to the weak-bond fracture of a lattice (Slepyan and Ayzenberg-Stepanenko, 2002), and to the transition waves in bistable structures (Slepyan and Troyankina (1984), Slepyan (2002), Slepyan et al (2005), Vainchtein (2010)). In a sense, the spontaneous crack propagation considered below also relates to the so called Prince Rupert's drop phenomenon of disintegration known from 17th century (see, e.g., Johnson and Chandrasekar, 1992) and to the fragmentation of metastable glass (Silverman et al, 2012). These phenomena can be referred to the internal potential energy, which releases under a local breakage, resulting in the spontaneous  disintegration.

The state of the considered structure is characterised by the following values. The first is the {\em initial internal energy}, $\CE$, which arises due to the difference in the interface bond lengths and stored in the cell of periodicity, that is, in two spans of the structure.  Next is the {\em initial energy of the bond}, $E$, which is the same for all the bonds. These two values correspond to the initial state of the intact structure. Finally, it is the {\em actual energy of the bond}, $E_m$, where $m$ is the bond number. These energies refer to the microlevel.

The critical values of the initial internal energy and the bond energy are denoted by $\CE_c$ and $E_c$, \res. The latter is the same for all the bonds; it relates to the critical extension. Note that $E=E_c$ under $\CE=\CE_c$. The level of the internal energy is characterised by the {\em ratio of the stored internal energy to its critical value}, $\Gg=\CE/\CE_c=E/E_c$.

In the formulation, the displacements about the initial positions of unstrained bond ends are introduced as ones in the initial and actual states, $\CU_m$ and  $u_m=\CU_m+U_m(t)$, \res, where $m$ is the bond number. It is assumed that the bond stiffnesses are the same. In this case, in the initial state
 \beq \CU_{2m} =- \CU_{2m+1} = \mbox{const}\,,~~~m=0, \pm1, ...\,.\eeq{00}

In the analytical study, we consider the bridge crack as the initially stretched bond breakage propagating in the {\em steady-state} mode. We use this term assuming that the dynamic state is characterised by two functions of one argument, $\Gn=t-\Gj m$, but not on $t$ and $m$ separately (a constant crack speed $v=1/\Gj$).  Numerical simulations demonstrate that the steady-state solutions found analytically really exist. In addition, the simulations reveal more complicated ordered modes with crack speed oscillations, where periodic clusters consist of more than one breaking bonds.

For the specified structures we determine the lower values of $\Gg=\Gg_c$ as a function of an orthotropy parameter, $\Ga$, as the lower bound of the domain where the bridge crack may propagate spontaneously. As a manifestation of the dynamic amplification factor (see Slepyan (2000)), this bound appears considerably below its static values found in Mishuris and Slepyan (2013). Next we determine the crack speed as a function of $\Gg, \Gg_c\le \Gg<1$. In the numerical simulations, it appears that the steady-state regime is quickly established if parameter $\Gg$ is not too close to $\Gg_c$. Otherwise, the crack speed is unstable and can be very low. It becomes supersonic and grows unboundedly for $\Gg\to 1$. Moreover, in this case, not only initially stretched bonds break as assumed in the analytical formulation, but the initially compressed bonds become broken, and an open propagating crack forms with a finite bridged region.

\section{Analytical study of a general problem}
\subsection{Problem formulation}

\begin{figure}[h!]
  %\begin{center}
    \hspace{15mm}\includegraphics [scale=0.6]{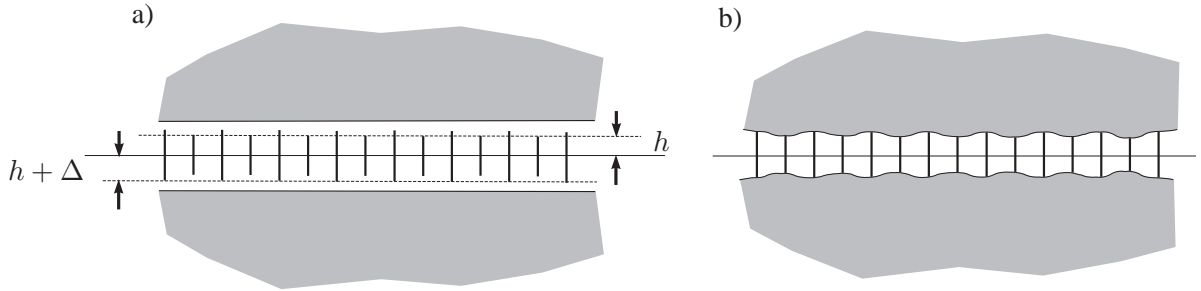}
    \put(-205,53){\small$h$}
    \put(-449,42){\small$h+\Delta$}
  %\end{center}
    \caption{\small The body with the structured interface. The compressed and stretched bonds alternate creating self-equilibrated stresses.}
\label{f11}
\end{figure}
We consider two equal half-planes or layers of a non-specified periodic structure connected by a set of elastic bonds, \fig{f11}b.
The  bonds are numbered by $m = 0, \pm 1, ...$ .  The between-the-bond distance is taken as the length unit. The even and odd bonds differ only by their initial length, namely, the even bonds, $m=0, \pm 2, ...,$ are of the length $2h$, whereas the odd bonds, $m=\pm 1, \pm 3, ...,$ are of a slightly different length, $2h+2\GD$ with $\GD>0, \GD/h\ll 1$. In the framework of the  Hooke's law, the bond's stiffness, $\Gk$, is assumed to be the same for both the even and the odd bonds, and the response of the structure to external forces corresponds to the regular, periodic set of the bonds. This also concerns the bulk of the body, where the internal energy may be present.

If the internal energy level is sufficient, a spontaneous breakage of the even bonds can propagate along the interface in the absence of external forces. We assume that the steady-state regime can exist, where the dynamic displacements of the upper knots of the interface, additional to those in the initial state, are described by two different functions of a single argument, one function for the even bonds and the other for the odd ones
\beq U_m(t) = U(\Gn) ~~~(m=0, \pm 2, ...)\,,~~~U_m(t) = V(\Gn)  ~~~(m=\pm 1, \pm 3, ...)\,,~~~\Gn=t-\Gj m\,.\eeq{new1}
Note that, in these terms, the bridged crack speed $v=1/\Gj$.

The structure's dynamic properties are reflected by the Green's function considered below. The initial displacements of the upper knots of the interface are defined by the static solution presented in Mishuris and Slepyan (2013). The additional displacements satisfy the relations \eq{new1}. No even bonds exist at $\Gn>0$ (it is the bridged crack area), while the structure is intact at $\Gn\le 0$.  The conditions at infinity correspond to the absence of external actions. In the analytical calculations, the crack speed  is an input parameter. In doing so, we determine the even bond state at the fracture point, $\Gn=0$, also as a function of $v$.

Recall that the inverse, multi-valued function is usually uniquely defined by the admissibility condition. Namely, the bond strain energy before the fracture point must be below the critical value. It follows from this that among the speeds, which are found to correspond to a given level of the internal energy, only the maximal speed is acceptable (Marder and Gross (1995)) and that the results for $\Gg(v)>1$ are not acceptable. Also, the results obtained under this formulation are valid if the odd-number bond dynamic states remain subcritical. Note that the numerical simulations discussed in \az{nss} confirm the analytical results for almost the whole range of $\Gg$, whereas it appears that, in the case where $\Gg$ is close to one, the odd bonds become broken too.

Mathematically, we consider the intact structure under self-equilibrated external forces, $Q(\Gn)$, acting in the opposite directions on the upper and lower ends of the even bonds. These forces, which are unknown in advance, serve to compensate the tensile forces in the even bonds at $\Gn>0$
 \beq Q(\Gn)=Q_0(\Gn)=  2\Gk u_0(\Gn)=2\Gk \gl[\CU_0 +U(\Gn)\gr]~~(m=0, \pm 2, ...)\,,\eeq{new0}
where $\Gk$ is the bond stiffness. The task is to determine the dynamic displacements, whereas the initial displacements found in Mishuris and Slepyan (2013) are
 \beq \CU_0=-\CU_1 =\GD L(0)\,,\eeq{cu01}
where $L(k)$ is the kernel of the static version of the Wiener-Hopf equation. Note that $L(0)$ takes the same value in both the dynamic and static formulations.

 \subsection{Dynamic Green's functions}
 In the analysis, we consider the dynamic crack-related Green's function,  $G(m,t)$, corresponding to the intact structure under unit self-equilibrated pulses acting at $t=0$ in opposite directions on the upper and lower ends of the bond $m=0$. It then follows that the dynamic displacements, $U(\Gn)$ and $V(\Gn)$, are
 \beq U(\Gn)= \sum_{m'=0,\pm 2, ...}G(m-m',t)\ast Q_0(t-\Gj m')~~~(m=0,\pm 2, ...)\,,\n
 V(\Gn)= \sum_{m'=0,\pm 2, ...}G(m-m',t)\ast Q_0(t-\Gj m')~~~(m=\pm 1, \pm 3, ...)
 \,,\eeq{new2}
where the asterisk means the convolution on $t$.

Below the Fourier transform on $t$ and the discrete Fourier transform on $m$ are denoted by the superscript $F$, whereas the continuous Fourier transform on $\Gn$ is denoted by the superscript $F_\Gn$. The latter leads to the following relations
 \beq
 U^{F_\Gn}(k)= \inti U(\Gn)\E^{\I k \Gn}\D \Gn  = Q_0^{F_\Gn}(k)\sum_{m'=0,\pm 2, ...}G^F(m-m',k)\E^{-\I k\Gj (m-m')}\,,\n
  V^{F_\Gn}(k) = \inti V(\Gn)\E^{\I k \Gn}\D \Gn =  Q_0^{F_\Gn}(k)\sum_{m'=0,\pm 2, ...}G^F(m-m',k)\E^{-\I k\Gj (m-m')}\eeq{UV1}
 with
 \beq Q_0^{F_\Gn}(k)=\int_0^\infty Q_0(t)\E^{\I k \Gn}\D \Gn\,.\eeq{Q0tF}
Thus
 \beq U^{F_\Gn}(k)= Q_0^{F_\Gn}(k)\CG_{even}(k) \,,~~~
 V^{F_\Gn}(k)= Q_0^{F_\Gn}(k)\CG_{odd}(k)\,.\eeq{UV2}
The functions $\CG_{even}(k)$ and $\CG_{odd}(k)$ are the double Fourier transforms, the continuous transform on $t$ with the parameter $k$ and the selective discrete transforms on $m$ (on even $m$ and on odd $m$ separately) with the parameter $-\Gj k$
 \beq \CG_{even}(k)= G_{even}^{FF}(-\Gj k,k)= \sum_{m=0,\pm 2, ...}\inti G(m,t)\E^{\I k t -\I \Gj k m}\D t\,,\n
 \CG_{odd}(k) =  G_{odd}^{FF}(-\Gj k,k)= \sum_{m=\pm 1,\pm 3, ...}\inti G(m,t)\E^{\I k t -\I \Gj k m}\D t\,.\eeq{new33}
The selective transforms introduced in Mishuris and Slepyan (2013) can be expressed in terms of the regular discrete transform as follows:
 \beq \CG_{even}(k)= G_{even}^{FF}(-\Gj k,k)=\f{1}{2}\gl[G^{FF}(-\Gj k,k) +G^{FF}(-\Gj k+\pi,k)\gr]\,,\n
 \CG_{odd}(k) =  G_{odd}^{FF}(-\Gj k,k)=\f{1}{2}\gl[G^{FF}(-\Gj k,k) - G^{FF}(-\Gj k+\pi,k)\gr]\,,\eeq{new5}
with
 \beq G^{FF}(-\Gj k,k) = \sum_{m=0,\pm 1, ...}\inti G(m,t)\E^{\I k t -\I \Gj k m}\D t\,.\eeq{new6}

\subsection{The spontaneous bridge crack propagation}
Let the breakage of the even bonds propagate with constant speed, $v=1/\Gj >0$, and let the damaged area be placed at $\Gn = t-m\Gj >0$, while the intact bond area be at $\Gn \le 0$. We can consider this structure as completely intact but under the external forces, $Q_0(\Gn)$, which compensate the action of the even bonds at $\Gn>0$. These forces are defined in \eq{new0}. The Fourier transform leads to
 \beq Q_0^{F_\Gn}(k) =  2\Gk u_+(k)=2\Gk\big(U_+(k) +\CU_0/(0-\I k)\big)\,,\n
U^{F_\Gn}(k)= U_+(k)+U_-(k)\,,~~~U_\pm(k)=\inti U(\Gn)\E^{\I k \Gn}H(\pm \Gn)\D \Gn \,.\eeq{dr1}
Recall that $\CU_0=-\CU_1$ are the initial displacements \eq{cu01} caused by the microlevel stresses.

Now the governing equations follow from  \eq{UV2} as
 \beq U^{F_\Gn}(k)= 2\Gk \CG_{even}(k)\big(U_+(k) +\CU_0/(0-\I k)\big)\,,\n
  V^{F_\Gn}(k) =  2\Gk \CG_{odd}(k)\big(U_+(k) +\CU_0/(0-\I k)\big)\,.\eeq{dr2a}

The first of the equations in \eq{dr2a} is independent of the second. It is the Wiener-Hopf type equation with respect to the dynamic displacements of the even bonds
 \beq U_-(k) + L(k)U_+(k) =  \gl[1-L(k)\gr]\CU_0/(0-\I k)\,,~~~ L(k)=1-2\Gk \CG_{even}^{F}(k)\,,\eeq{drgf41}
where $\CG_{even}(k)$ is expressed through the original Green's function $G^{FF}(-\Gj k,k)$ in \eq{new5}.

We assume that the function $L(k)$ satisfies the conditions which allow us to solve this equation in a regular way (in this connection, see Slepyan, 2002, pp 449-451). To proceed, the factorization should be made as
 \beq L(k)=\lim_{\Im k\to 0}L_+(k)L_-(k)\,,~~~L_\pm(k) = \exp\gl[\pm\f{1}{2\pi\I}\inti \f{\ln L(\xi)}{\xi- k}\D \xi\gr]~~~(\pm\Im k >0)\,.\eeq{fact1}
In particular, it follows from this that
\beq L_\pm(\pm \I\infty) =1\,,~~~ L_\pm(0) = \sqrt{L(0)}\,\CR^{\pm 1}\,,~~~\CR=\exp\gl[\f{1}{\pi}\int_0^\infty \f{\mbox{Arg} L(\xi)}{\xi}\D \xi\gr] \,.\eeq{fact1a}

Next, we represent Eq. \eq{drgf41} in a form, where the plus/minus functions are separated
 \beq \f{U_-(k)}{L_-(k)} +L_+(k)U_+(k) =\f{\CU_0}{0-\I k}\gl[\f{1}{L_-(k)} -L_+(k)\gr]=C_+(k)+C_-(k)\,,\n
 C_+(k) = \f{\CU_0}{0-\I k}\gl[\f{1}{L_-(0)} -L_+(k)\gr]\,,~~~C_-(k)=  \f{\CU_0}{0-\I k}\gl[\f{1}{L_-(k)} -\f{1}{L_-(0)}\gr]\,.\eeq{fac2}
In the considered case, it follows that
 \beq U_+(k) =  \f{\CU_0}{0-\I k}\gl[\f{1}{L_-(0)L_+(k)}-1\gr]\,,~~~U_-(k) =  \f{\CU_0}{0-\I k}\gl[1-\f{L_-(k)}{L_-(0)}\gr]\eeq{fac3}
and
 \beq U(0) = \lim_{k\to -\I\infty}\I k U_-(k)= \lim_{k\to \I\infty}(-\I k)U_+(k)= \CU_0\big(1/L_-(0)-1\big)\,,\n
 u_0(0) = \CU_0/L_-(0)= \CU_0 \CR/\sqrt{L(0)}\,.\eeq{fac4}
To calculate this value for a specific structure the integral transform of the original Green's function, $G^{FF}(-\Gj k,k)$, should be specified. As for $V(\Gn)$, its transform, $V^{F}(k)$, is defined by the last relation in \eq{dr2a}, where $U_+(k)$ is now known \eq{fac3}.

\subsection{Energy relations}
The critical strain energy, $E_c$, is crack-speed independent
 \beq E_{c}= 2\Gk u_c^2\,,\eeq{er1}
where $u_c$ is the critical displacement. The internal energy density per two span, $\CE$, is (see Mishuris and Slepyan (2013))
 \beq \CE = 2\Gk\GD^2L(0)\,.\eeq{ss5}

We introduce the ratio
 \beq \Gg = \CE/\CE_c=\CU_0^2/u_c^2\,,\eeq{cece1}
where $\CE_c$ is the critical internal energy: $\CE=\CE_c$ as $E=E_c$ in the initial state. At the moment of the bond breakage the total displacement $\CU_0+U(0)=u_c$, and it follows from \eq{fac4} that
 \beq \Gg = L_-^2(0) = L(0)\CR^{-2}\,.\eeq{gammmm}
Note that the right-hand side of \eq{gammmm} is a function of the crack speed. So this relation serves for the determination of the crack speed as a function of $\Gg$. Recall that \eq{gammmm} may be satisfied by a number values of $v$. Usually (Marder and Gross, 1995) only the maximal value of the speed is considered to be admissible, because it corresponds to the first moment when the bond state becomes critical. There are some exceptions, however, as we show below.

The fracture energy itself  is found based on \eq{fac4}, \eq{ss5} and \eq{cu01}. It is
 \beq E_f = 2\Gk (\CU_0+U(0))^2 = 2\Gk \CU_0^2L_-^{-2}(0)\CR^2 = 2\Gk\GD^2L(0)\CR^2\,.\eeq{cfe1}
Thus, the ratio of the fracture  energy to the internal energy, $\CE$, is
 \beq R_f=\f{E_f}{\CE}=\CR^2=\exp\gl[\f{2}{\pi}\int_0^\infty \f{\mbox{Arg} L(\xi)}{\xi}\D \xi\gr]= \f{L(0)}{\Gg}\,.\eeq{errrf}

\section{Two specific structures}
Here we consider the same specific structures as in Mishuris and Slepyan (2013): the two-line chain and the orthotropic square lattice, but we will take into account point masses placed at the knots. The spontaneous failure waves in these structures demonstrate a nontrivial dependence on the internal energy level and the anisotropy parameter. Below we illustrate and discuss this in detail.

\subsection{The two-line chain}

\begin{figure}[h!]
    \hspace{6mm}
    \begin{picture}(0,0)(-10,45)
    \includegraphics [scale=0.50]{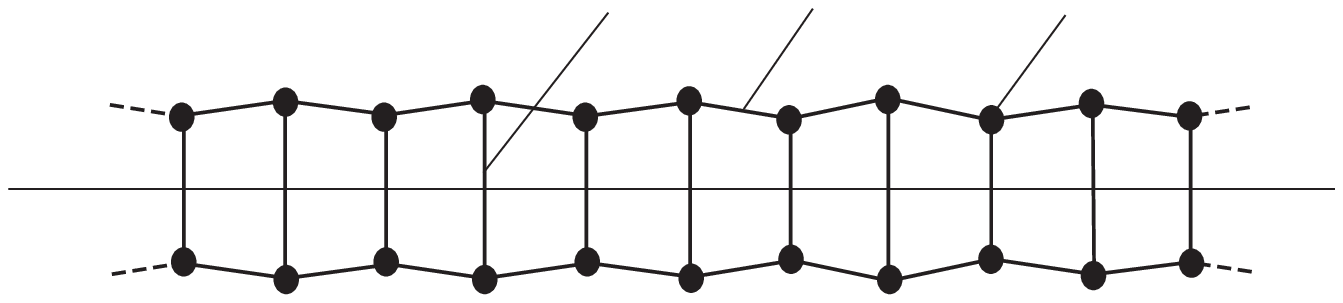}\hspace{10mm}
    \put(17,-13){\small$m=-6$\hspace{0.5mm}$-5$\hspace{0.5mm}$-4$\hspace{0.2mm}$-3$%
    \hspace{0.1mm}$-2$\hspace{0.5mm}$-1$\hspace{3mm}$0$\hspace{3.2mm}$1$\hspace{3.3mm}$2$\hspace{3.3mm}$3$\hspace{3.2mm}$4$}
    \put(20,46){\small b)}
    \end{picture}

    %\vspace{-60mm}

    \begin{picture}(0,60)(-250,-15)
  \includegraphics [scale=0.50]{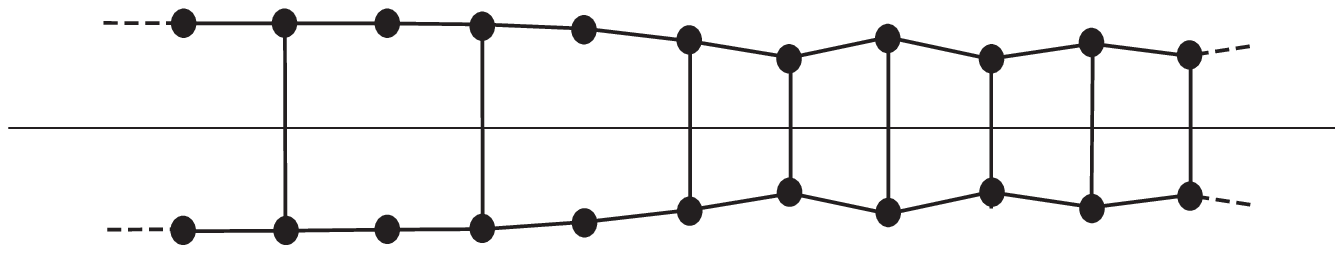}
    \put(-421,-13){\small$m=-4$\hspace{0.5mm}$-3$\hspace{0.5mm}$-2$\hspace{0.5mm}$-1$%
    \hspace{2.7mm}$0$\hspace{3mm}$1$\hspace{3mm}$2$\hspace{3.2mm}$3$\hspace{3.3mm}$4$\hspace{3.3mm}$5$\hspace{3.2mm}$6$}

\put(-415,46){\small a)}
\put(-321,46){\small$\varkappa$}
\put(-290,46){\small$\mu$}
\put(-256,46){\small$M$}

\end{picture}
    \caption{\small The intact chain (a) and the chain with a semi-infinite bridged crack (b). }
\label{ff2}
\end{figure}

The dynamics of the chain shown in \fig{ff2} is governed by the equation
 \beq M\ddot{u}_m(t) +2(\Gk +\Gm)u_m(t) - \Gm (u_{m+1}(t)+u_{m-1}(t))=\pm Q(m,t)\,,\eeq{detlc1}
where signs $\pm$ correspond for the upper and lower line, \res.
From this, the Green's function $G(m,t)$ is defined as the displacements corresponding to $Q(m,t) = \Gd(t)\Gd_{m,0}$. We find
 \beq G^{FF}(-\Gj k,k)= (1/\Gm)[(0-\I k/c)^2 +2(\Ga+1-\cos \Gj k)]^{-1}\,,~~~c=\sqrt{\Gm/M}\,.\eeq{detlc2}
It follows that
 \beq \CG_{even}=\f{1}{\Gm}\f{(0-\I k/c)^2 + 2(\Ga+1)}{[(0-\I k/c)^2 + 2(\Ga+1)]^2 - 4\cos^2 \Gj k}\,,\n
 \CG_{odd}=\f{1}{\Gm}\f{2\cos \Gj k}{[(0-\I k/c)^2 + 2(\Ga+1)]^2 - 4\cos^2 \Gj k}\,,\n
  L_1(k)=(0-\I k/c)^4+ 2(\Ga+2)(0-\I k/c)^2 + 4\Ga+4\sin^2\Gj k\,,\n
  L_2(k)= (0-\I k/c)^4 + 4(\Ga+1)(0-\I k/c)^2 +4\Ga(2+\Ga)+ 4\sin^2 \Gj k\,,\n
   L(k) =L_1(k)/L_2(k)\,,\quad
 L(0) = 1(2+\Ga)\,.\eeq{detlc3}

Note that the dimension of $\Gj, 1/k, 1/c$ is time. Along with this, since the between-the-bond distance is taken as the length unit, $c$ is the long wave speed. It can be seen below that the spontaneous failure wave can propagate with hypersonic speeds, $v=1/\Gj\gg c$.

\subsubsection{Some limiting relations}
We now derive some limiting relations which give us reference points for the dependencies presented below. Let us represent expressions corresponding to zero points of $L_{1,2}(k)$  in the form as
 \beq k_{1,2}^2/c^2= \Ga+2\mp\sqrt{\Ga^2+4\cos^2\Gj k_{1,2}}~~~(L_1(k_{1,2})=0)\,,\n
 k_{3,4}^2/c^2 = 2(\Ga+1)\mp 2|\cos\Gj k_{3,4}|~~~(L_2(k_{3,4})=0)\,.\eeq{zpol12}
Note that these points correspond to waves radiated during the bridge crack propagation. Namely, the numbers, $k_i, i=1,...,4$, are the wave frequencies and $\Gj k_i$ are the wavenumbers.

Based on \eq{detlc3} and \eq{zpol12} we can find limiting relations corresponding $\Ga=0, \Ga=\infty$ and $v=1/\Gj =\infty$. In the case $\Ga=0$
 \beq L_1(k)=L_2(k), ~~\mbox{Arg}L(k)=0\,,~~\CR=1\,,~~L(0)=1/2\,,\eeq{Ggo0}
and it follows from \eq{gammmm} that $\Gg$ takes the static value, $\Gg=1/2$.

The same conclusion is valid for the opposite case, $\Ga\to\infty$, and also for $v\to\infty~(\Gj\to 0)$. We have
 \beq k_{1,2}/c = \sqrt{\Ga + 2 \mp\Ga} +o(1)\,,~~~k_{3,4}/c=\sqrt{2\Ga} \mp \Ge_{1,2}\,,~~\Ge_{1,2}=o(1)~~~(\Ga\to\infty)\,,\n
 k_{1,2}/c =\sqrt{\Ga+2\mp\sqrt{\Ga^2+4}}\,,~~~k_{3,4}/c=\sqrt{2(\Ga+1)\mp 2}~~~(v=\infty)\,.\eeq{Ggo2}
 In the considered cases, Arg$L(k)$ is a piecewise constant function
\beq \mbox{Arg} L(k)=-\pi[H(k-k_1)+H(k-k_2)-H(k-k_3)-H(k-k_4)]\,,\eeq{HHHH}
and
\beq \CR=k_1k_2/(k_3k_4)\,.\eeq{RRR}
It follows that for any $v$
 \beq  \CR \sim 1/\sqrt{\Ga}\,,~~~\Gg \sim 1/\big(\Ga \CR^2\big)\to 1~~~(\Ga\to \infty)\eeq{RaRv1}
 and for any $\Ga$
\beq  \CR=1/\sqrt{\Ga+2}\,,~~~\Gg=1/\big((\Ga+2)\CR^2\big)=1~~~(v = \infty)\,.\eeq{RaRv}

Graphical  illustrations of analytical results obtained for the chain are presented in \az{ttlch} together with the results of numerical simulations.

\subsection{The square-cell lattice}\label{tscl}

\begin{figure}[h]
\begin{center}
\includegraphics[scale=0.50]{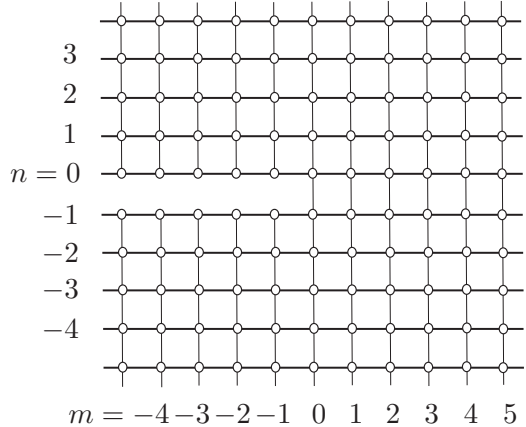}
\put(-173,-13){\small$m=-4$\hspace{0.5mm}$-3$\hspace{0.5mm}$-2$\hspace{0.5mm}$-1$%
    \hspace{2.7mm}$0$\hspace{3mm}$1$\hspace{3mm}$2$\hspace{3.2mm}$3$\hspace{3.3mm}$4$\hspace{3.3mm}$5$}
\put(-183,62){\small$-1$}\put(-195,78){\small$n=0$}\put(-175,93){\small$1$}\put(-175,108){\small$2$}\put(-175,123){\small$3$}
\put(-183,48){\small$-2$} \put(-183,34){\small$-3$}\put(-183,19){\small$-4$}
\end{center}
\caption{\small The orthotropic lattice. The stiffness of the horizontal and vertical bonds are $\Gm$ and $\Gk$, \res ($\Ga=\Gk/\Gm$). The bonds on the crack line are prestressed as defined above.}
\label{ff3}
\end{figure}

From the equation for the lattice dynamics under the unit self-equilibrated pulses
\[
 M\ddot{u}_{m,n}(t) +2(\Gk +\Gm)u_{m,n}(t) - \Gm (u_{m+1,n}(t)+u_{m-1,n}(t))-
\]
\beq
 \Gk(u_{m,n+1}(t)+u_{m,n-1}(t))=\Gd(t)\Gd_{m,0}[\Gd_{n,0}-\Gd_{n,-1}]\eeq{detlcscl1}
we find the double Fourier transform of the crack-related dynamic Green's function
 \beq G^{FF}(-\Gj k,k) = \f{1}{2\Gk}\gl(1-\sqrt{\f{Y +4\sin^2\Gj k/2}{Y +4\sin^2\Gj k/2+4\Ga}}\gr)\,,\n
 L(k) = \f{1}{2}\gl(\sqrt{\f{Y +4\sin^2\Gj k/2}{Y +4\sin^2\Gj k/2+4\Ga}}+\sqrt{\f{Y +4\cos^2\Gj k/2}{Y +4\cos^2\Gj k/2+4\Ga}}\gr)\,,\n
 L(0) = 1/\big(2\sqrt{\Ga +1}\big)\,,~~~Y=(0-\I k/c)^2\,,~~~c=\sqrt{\Gm/M}
 \,.\eeq{detlcscl3}
Note that such a lattice in absence of the  internal energy was considered in Mishuris at al (2007) without evaluation of the Green's function.

The limiting relations following from \eq{gammmm} and \eq{detlcscl3} are
\beq
\Gg(v/c,\alpha) \to 1/2~~(\Ga\to 0)\,,\n
\Gg(v/c,\alpha) \to \Gg_\infty(v/c)=\f{1}{2}\exp\gl[-\f{2}{\pi}\int_0^2\f{\mbox{Arg} L^*(k)}{k}\D k\gr]~~(\Ga\to\infty)\,,
\eeq{crit_1/2}
where
 \beq
 L^*(k) = \sqrt{(0-\I k)^2 +2(1-\cos \Gj c k)} +\sqrt{(0-\I k)^2 +2(1+\cos \Gj c k)}\,.\eeq{gaminf}
It is found from this that in the limit, $\Ga\to\infty$, the minimal value of $\Gg$ and the corresponding speed are $\Gg\approx 0.66656,\, v/c \approx 0.709$ (see \fig{f111}a, where $\Gg_{min}(\Ga)$ is plotted) .

\subsubsection{Graphical illustrations of the analytical results for the lattice}
\begin{figure}[h!]
  \begin{center}
    \includegraphics [scale=0.38]{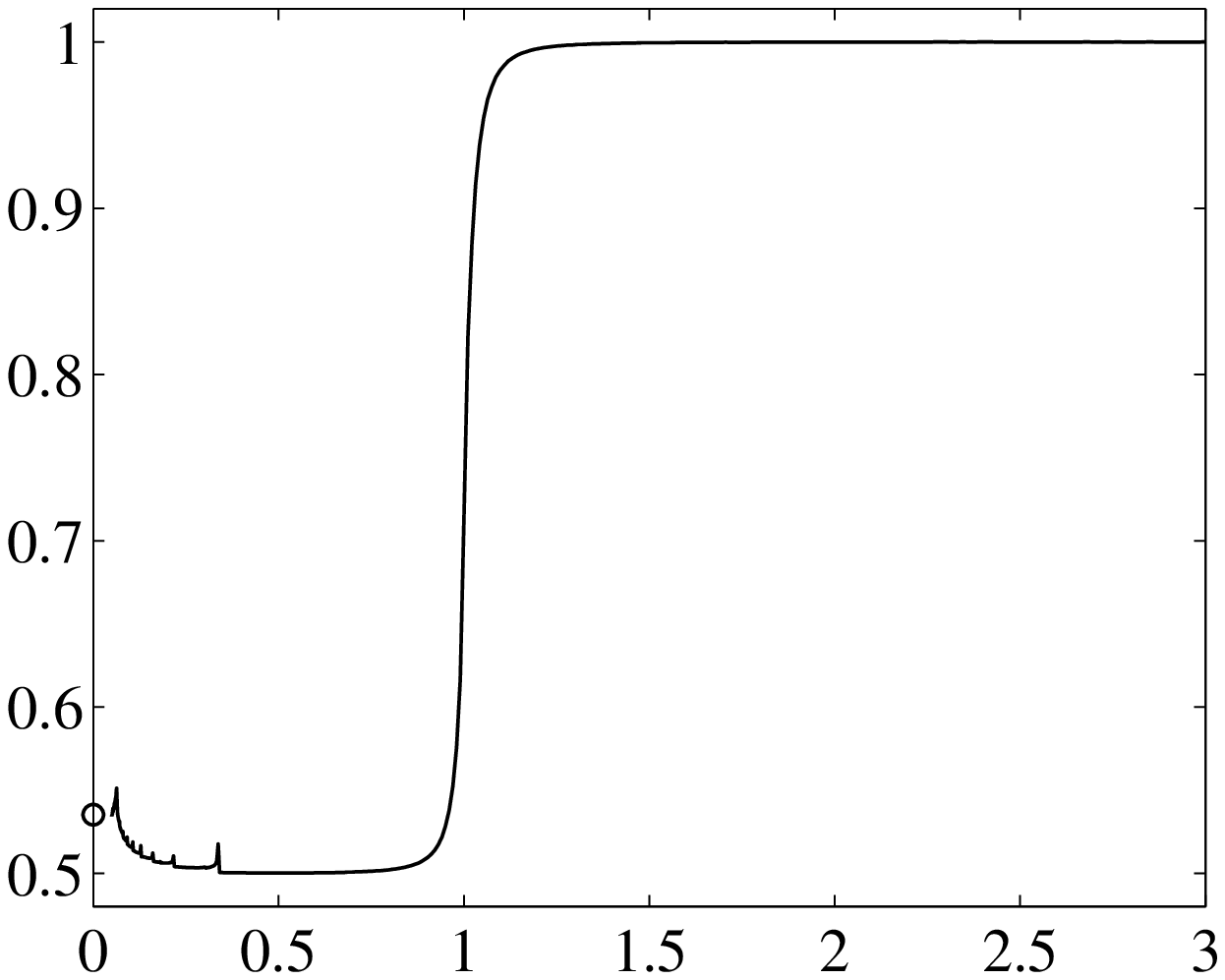}\includegraphics [scale=0.38]{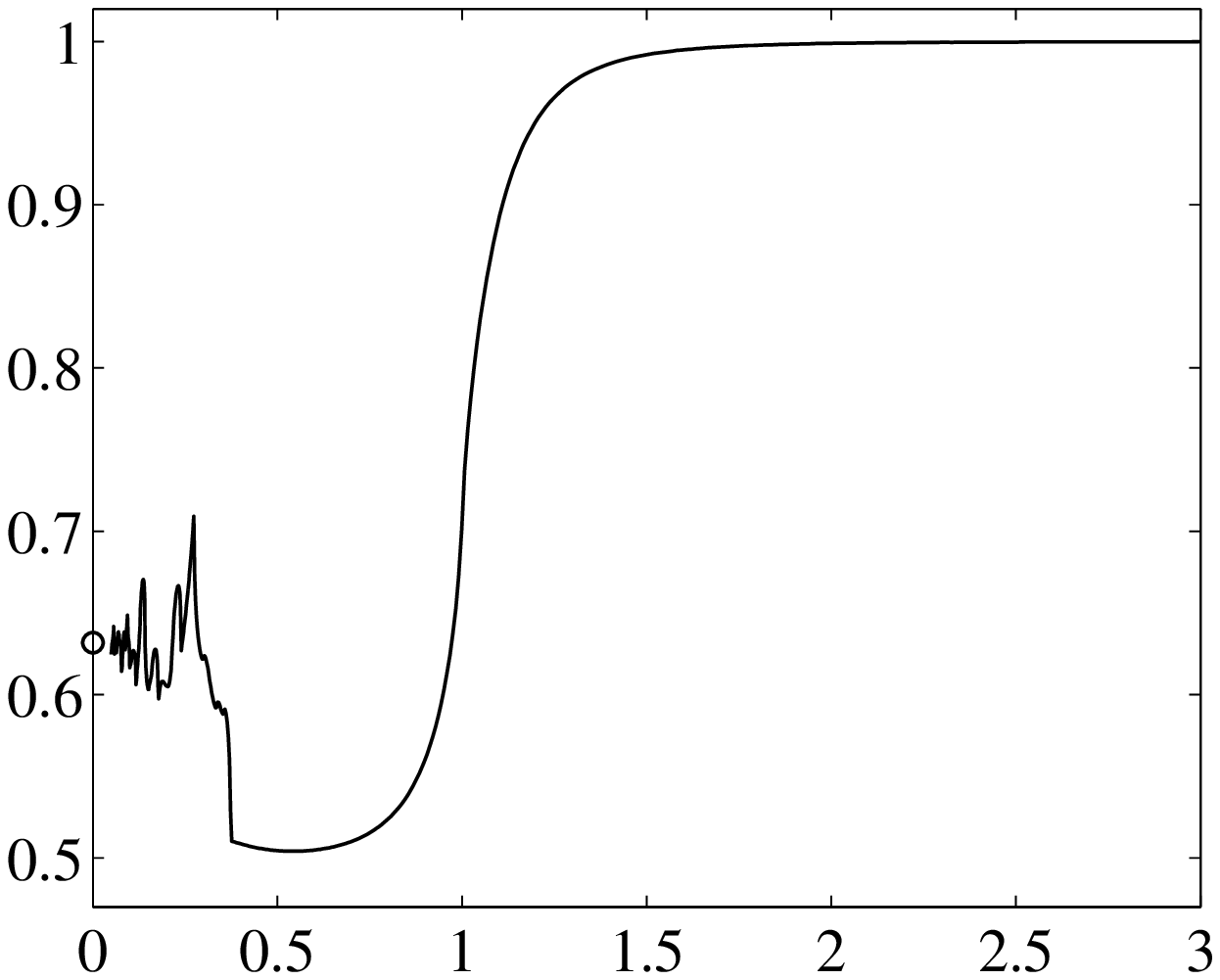}\includegraphics [scale=0.38]{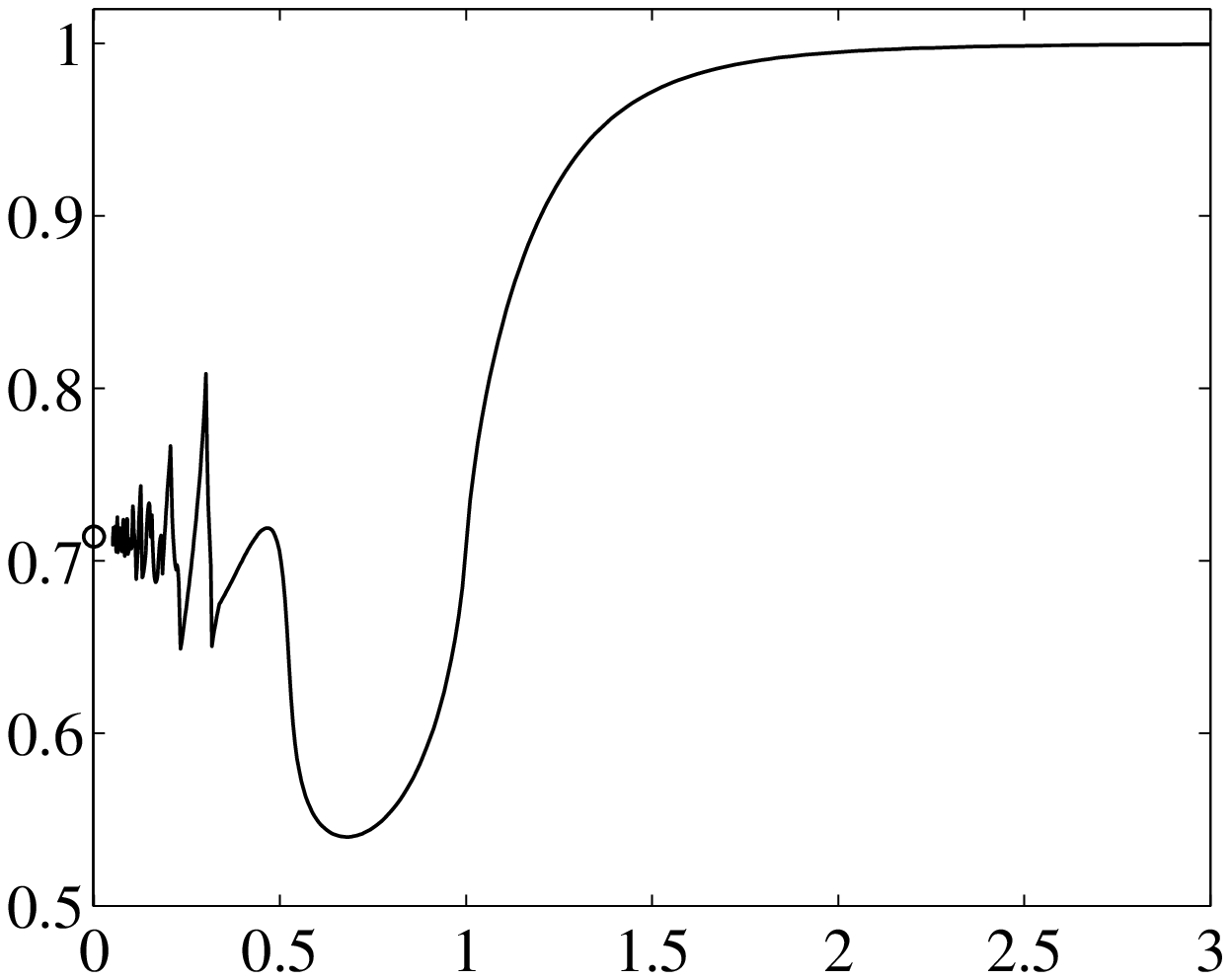}
       \put(-410,-5){\small$v/c$}\put(-250,-5){\small$v/c$}\put(-90,-5){\small$v/c$}
       \put(-480,80){$\gamma$}\put(-320,80){$\gamma$}\put(-160,80){$\gamma$}
        \put(-385,20){\small$\alpha=0.01$}\put(-225,20){\small$\alpha=0.2$}\put(-55,20){\small$\alpha=1$}
         \put(-480,120){a)}\put(-320,120){b)}\put(-160,120){c)}

  \end{center}
\caption{\small The spontaneous bridge crack in the lattice. The evolution of the $\Gg - v/c$ dependence when changing the anisotropy parameter, $\Ga$. Recall that $\Gg=\CE/\CE_c$ is the ratio of the internal energy to its critical value. The blurred minimum, where an unsteady crack speed regime can be expected, and a jump-like dependence corresponding to small $\Ga$. The minimum becomes more localized as $\Ga$ grows. }
\label{small_lattice}
\end{figure}

\begin{figure}[h!]
  \begin{center}
    \includegraphics [scale=0.38]{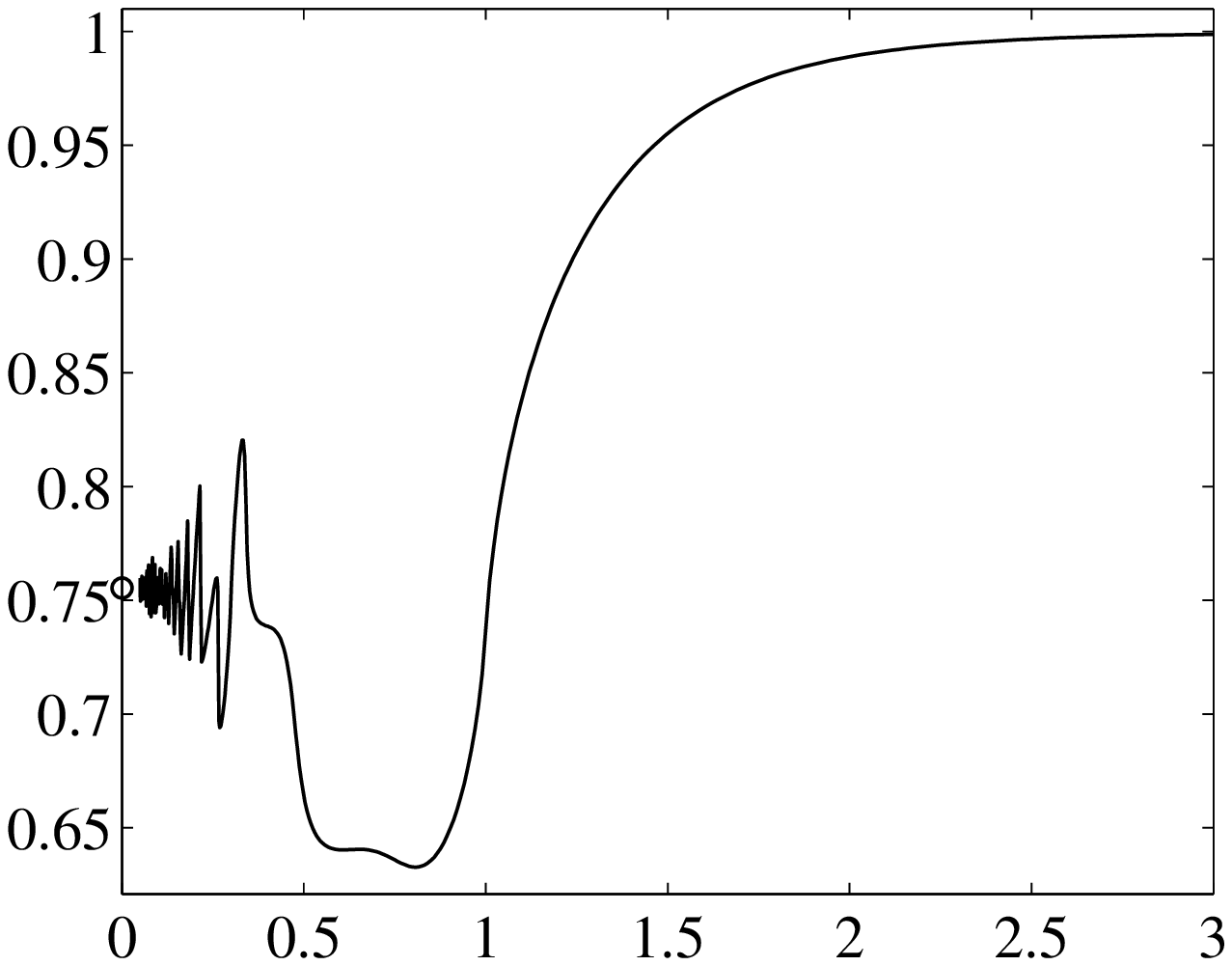}\includegraphics [scale=0.38]{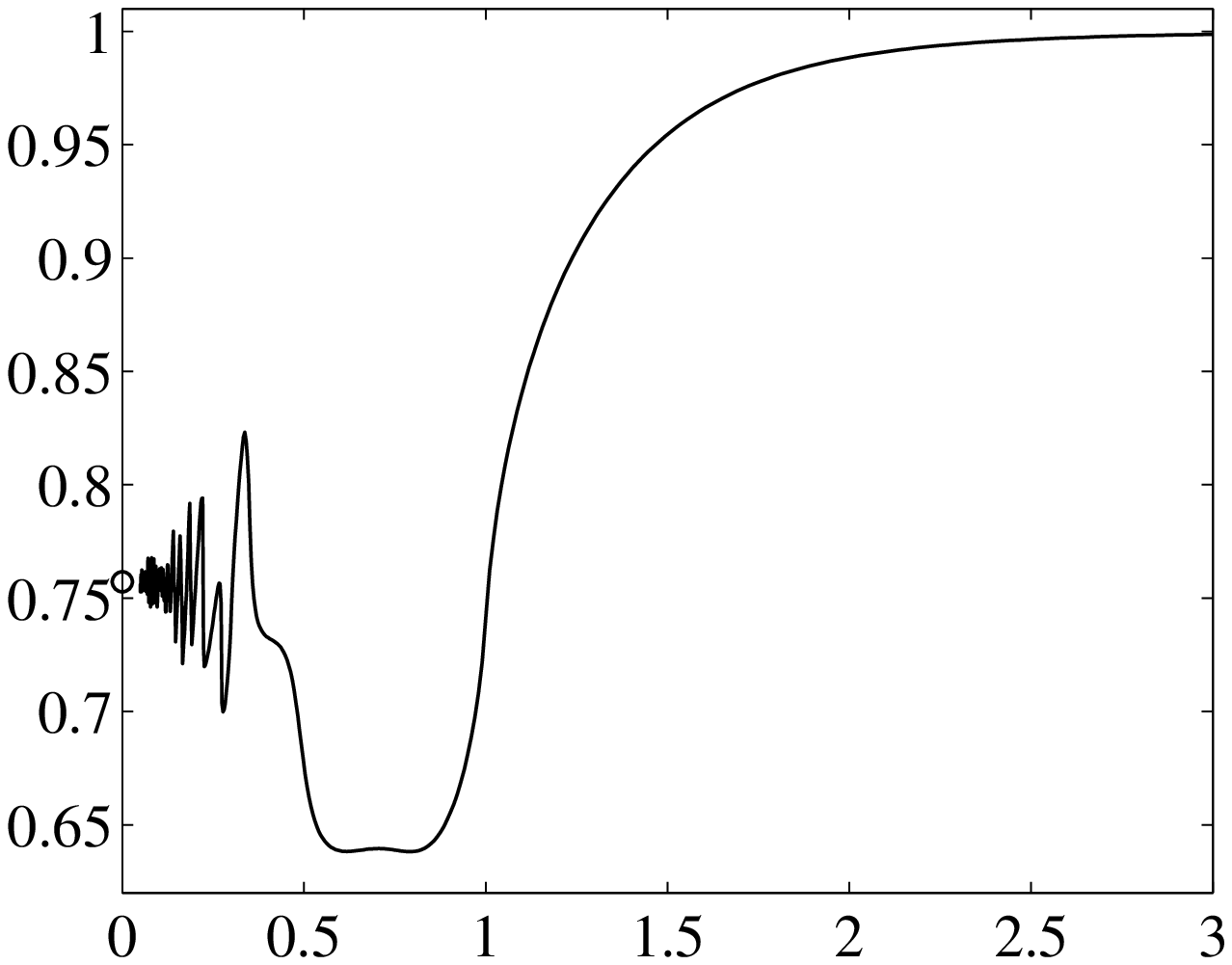}\includegraphics [scale=0.38]{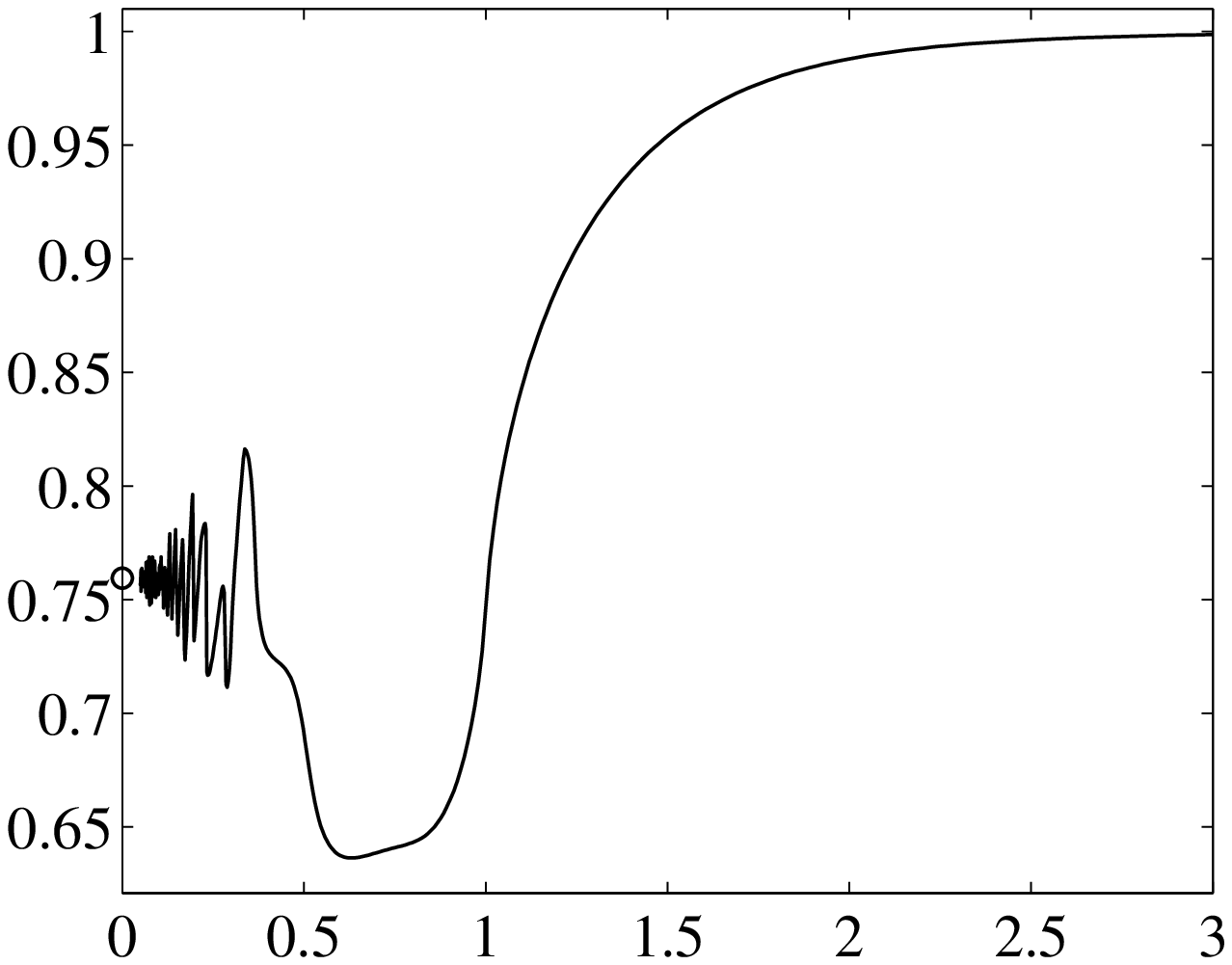}
       \put(-410,-5){\small$v/c$}\put(-250,-5){\small$v/c$}\put(-90,-5){\small$v/c$}
       \put(-480,80){$\gamma$}\put(-320,80){$\gamma$}\put(-160,80){$\gamma$}
        \put(-375,20){\small$\alpha=3$}\put(-230,20){\small$\alpha=3.215$}\put(-60,20){\small$\alpha=3.5$}
         \put(-480,120){a)}\put(-320,120){b)}\put(-160,120){c)}

  \end{center}
    \caption{\small The spontaneous bridge crack in the lattice. The evolution of the $\Gg - v/c$ dependence with the anisotropy parameter, $\Ga$. In these plots, one can see how the minimum bifurcates and changes the location under a relatively small change of the anisotropy parameter $\Ga$. The presence of two concurrent minima ($\Ga=3.215$) may lead to the crack speed oscillations about a stable averaged value.}
\label{f110}
\end{figure}

In this section, we present energy-speed relation plotted for some values of $\Ga$ in \fig{small_lattice} and \fig{f110} in the form of the dependence of the normalised internal energy, $\Gg=\CE/\CE_c$, on the normalized crack speed, $v/c$.

It can be seen that the minimal admissible speed is about a half of the long wave speed, $c$. So there is no slow crack in this case. This is the same as for the open crack in the isotropic lattice under remote forces (Slepyan, 1981). However, in contrast to the latter case, the crack in the lattice with the internal energy can propagate at any high speed, the long wave speed is not the crack speed limit anymore. Indeed, the crack gets the energy from the source distributed over its path but not from any forces acting at a distance.

In \fig{small_lattice} and \fig{f110}, a nontrivial role of the anisotropy can be seen: the blurred minimum, where an unsteady crack speed regime can be expected, a jump-like dependence corresponding to small $\Ga$ and the localization of the minimum with growing $\Ga$, \fig{small_lattice}.
In the range $3\le \Ga\le 3.5$, the minimum bifurcates and changes the location, and we meet two equal minima at $\Ga=3.215$, \fig{f110}. Uncertainty of the crack speed arises in this case. As discussed in \az{ttlch}, such a configuration results in the crack speed oscillations between these minima. This phenomenon resembles the clustering revealed earlier (see Mishuris et al, 2009, and Slepyan et al, 2010).

\begin{figure}[h!]
  \begin{center}
    \includegraphics [scale=0.38]{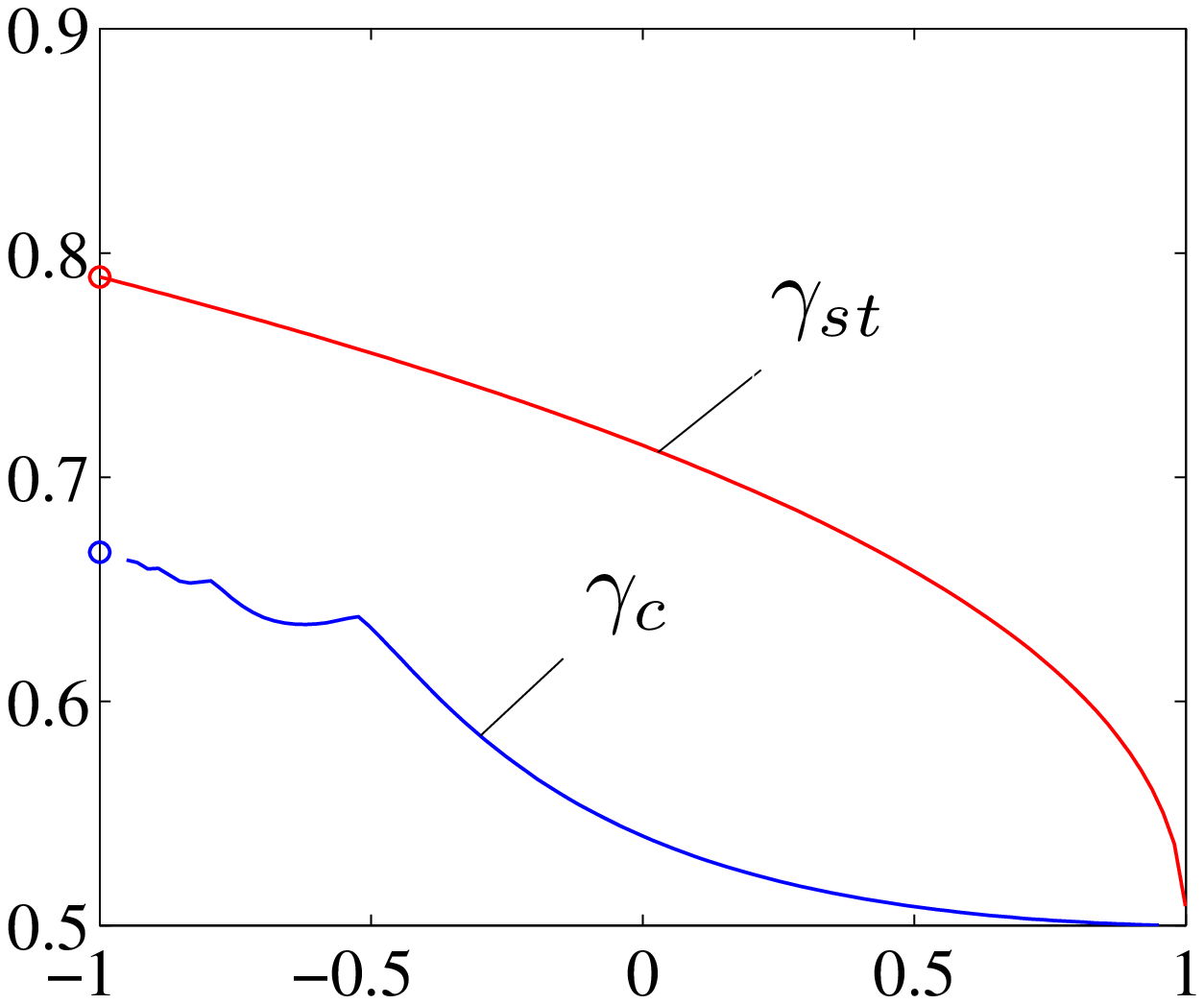}\includegraphics [scale=0.39]{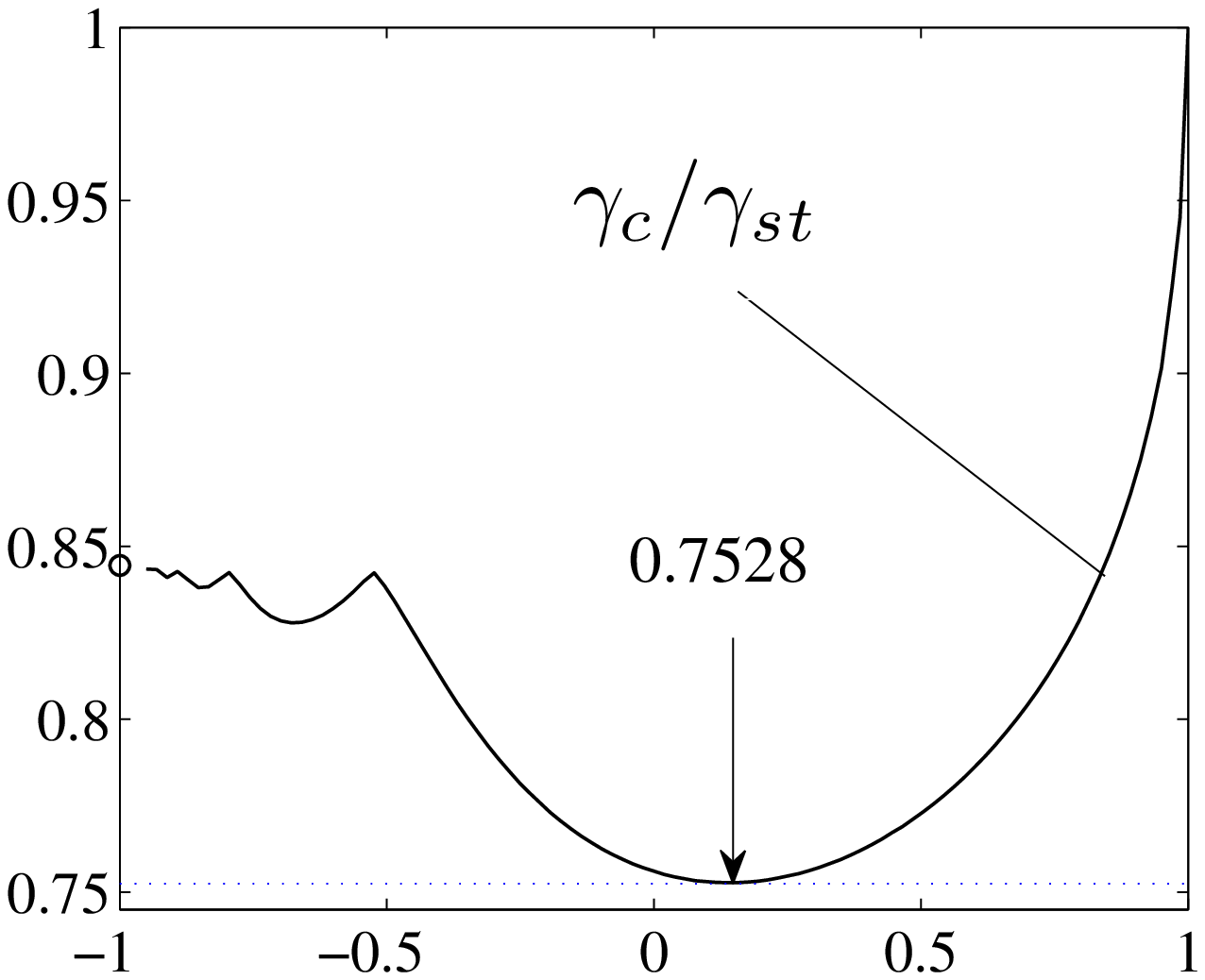}\includegraphics [scale=0.38]{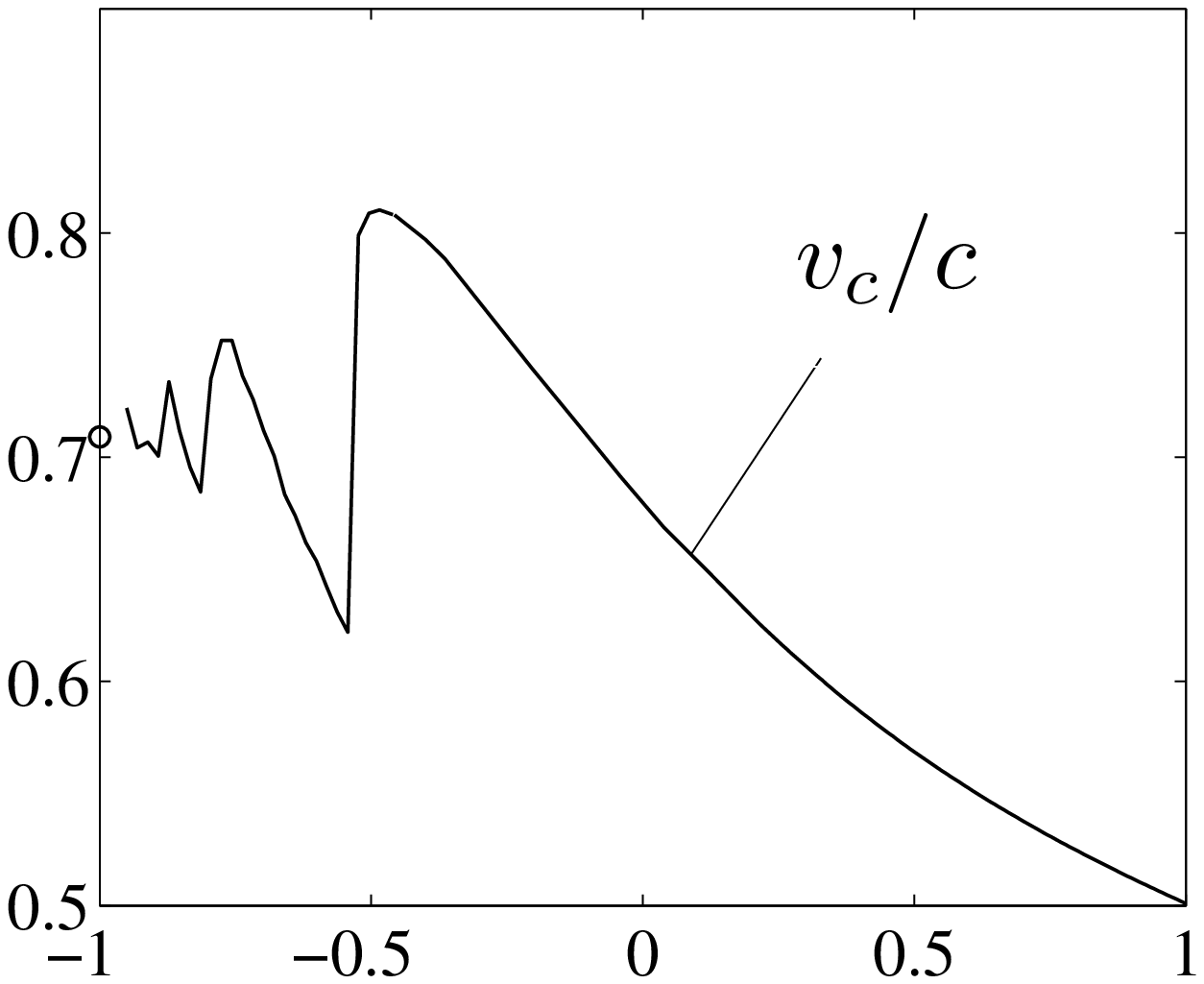}
      \put(-410,-5){\small$\hat\alpha$} \put(-245,-5){\small$\hat\alpha$}\put(-80,-5){\small$\hat\alpha$}
       \put(-480,130){a)}\put(-320,130){b)}\put(-160,130){c)}

  \end{center}
    \caption{\small The spontaneous bridge crack in the lattice: The dynamic, $\Gg_c$, and static, $\Gg_{st}$, values of the minimal $\Gg$, which is sufficient for the spontaneous crack to exist, as a function of the normalised anisotropy parameter, $\hat{\Ga}=(1-\Ga)/(1+\Ga)$ (a,b), and the corresponding crack speed (c). The small circles correspond to the limiting values found in \az{tscl}}
\label{f111}
\end{figure}

The minimal $\Gg$ as a function of the normalised anisotropy parameter, $\hat{\Ga}=(1-\Ga)/(1+\Ga)$, the dynamic and static values, and the corresponding crack speed are presented in \fig{f111}. Note that the minimal values of the internal energy required for the spontaneous crack growth initiation (in the static formulation) and propagation (the dynamic formulation) differ greatly. The decrease  of the latter in comparison with the former arises due to the dynamic factor (Slepyan, 2000). It is remarkable that there exists a jump in the dependence of the corresponding speed on $\hat{\Ga}$ at $v=0.5c$ (\fig{f111}c).

\section{The two line chain:  analytical and numerical results}\label{ttlch}
The main objectives of this paper are to determine the critical internal energy level, under which the spontaneous separation wave can exist, and to find the energy-dependent crack speed. At the same time, in the course of the analysis, we met some unexpected phenomena, such as: the crack speed oscillations at low internal energy, energy-dependent finite bridge zones at high energy level, discontinuity in the speed-energy dependence and a notable role of the structure anisotropy. We discuss all this by using the example of the two-line chain.

Note that, in the representation of the numerical results, $v$ is the current speed as before. Recall that for the fully bridged crack it is defined as $2/t_2$, where $t_2$ is the time between the breakage of two neighboring initially stretched bonds. The current speed corresponding to the other bond breakage, if it exists, is calculated similarly. In addition, in more complicated cases, the over-total-time-averaged speed is introduced as $\langle v\rangle= l(t)/t$, where $l(t)$ is the dynamic crack length, the fully/partially bridged crack length or the open crack length, \res.

\begin{figure}[h!]
  \begin{center}
    \includegraphics [scale=0.37]{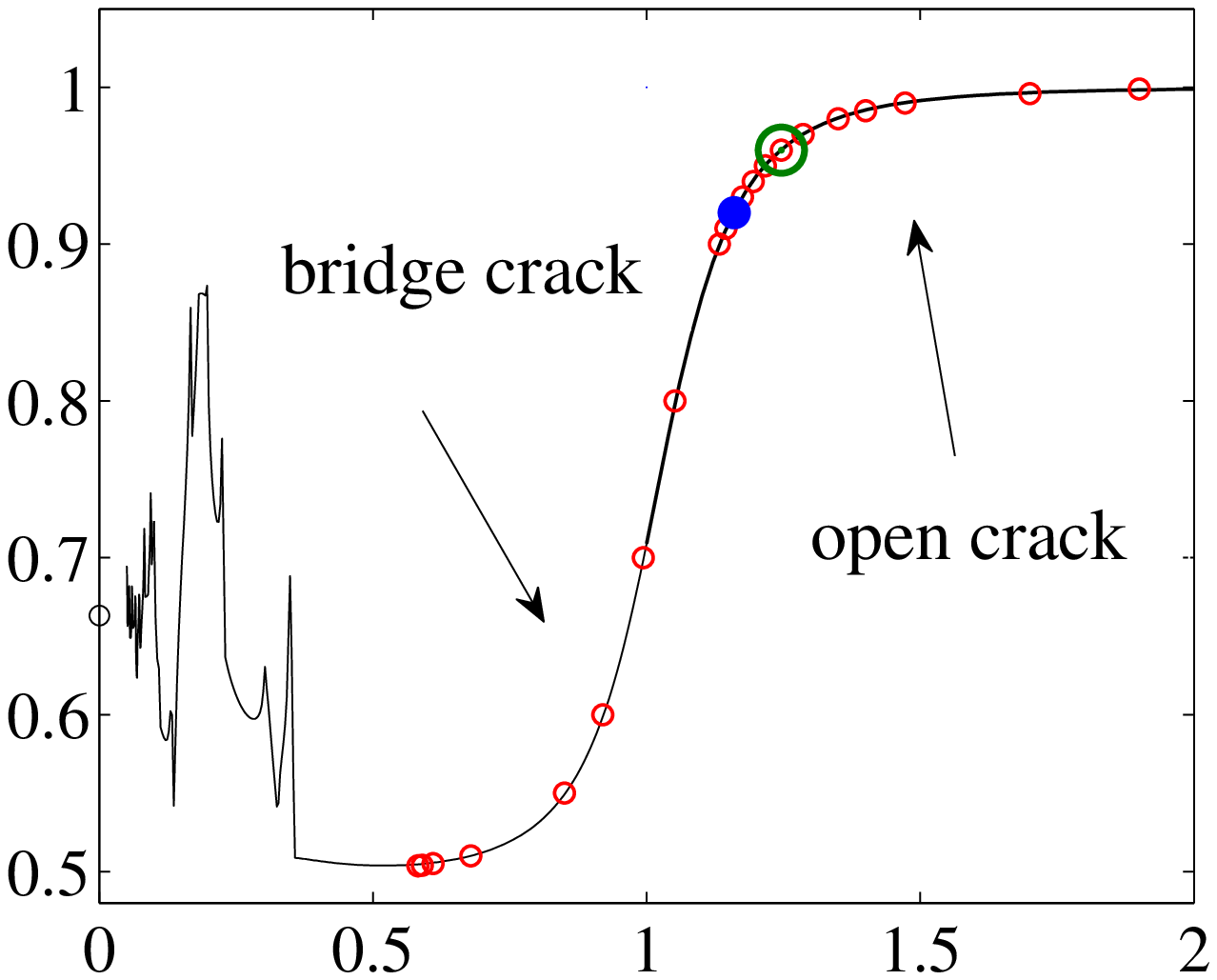}\includegraphics [scale=0.37]{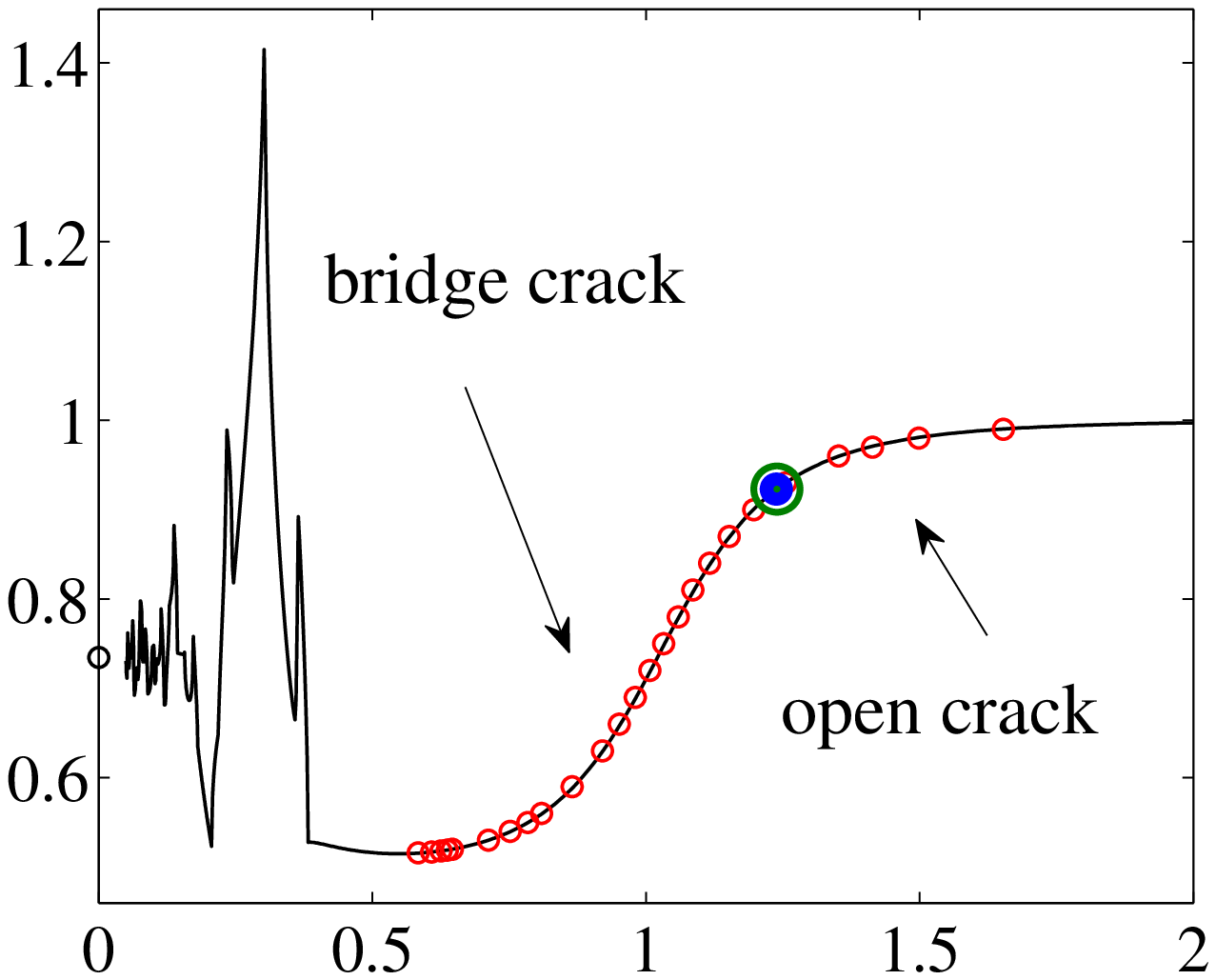}\includegraphics [scale=0.37]{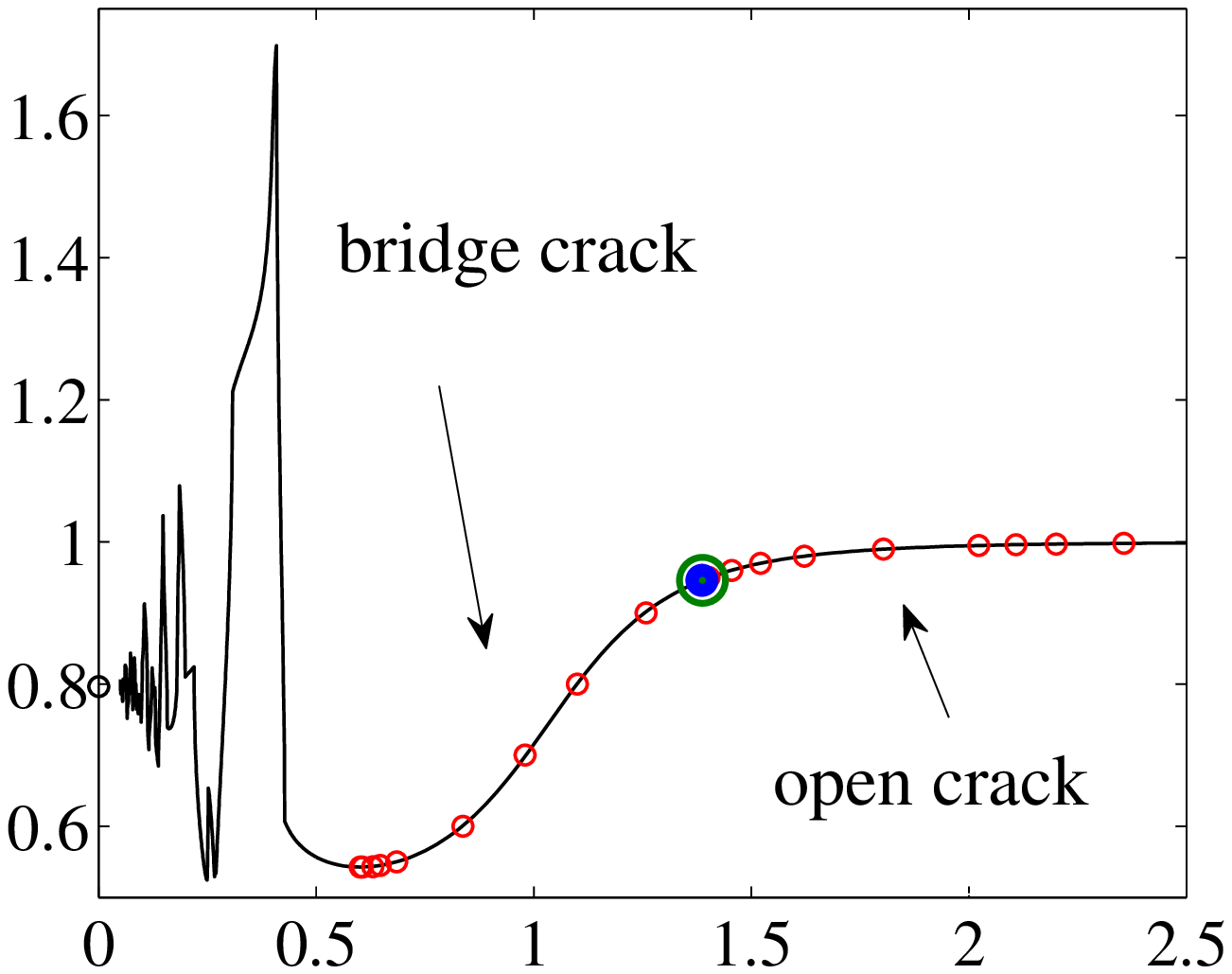}
       \put(-410,-5){\small$v/c$}\put(-250,-5){\small$v/c$}\put(-90,-5){\small$v/c$}
       \put(-482,80){$\gamma$}\put(-321,80){$\gamma$}\put(-161,80){$\gamma$}
        \put(-450,100){\small$\alpha=0.2$}\put(-220,100){\small$\alpha=0.5$}\put(-50,100){\small$\alpha=1$}
       \put(-480,120){a)}\put(-320,120){b)}\put(-160,120){c)}

  \end{center}
    \caption{\small  The spontaneous semi-infinite bridged crack in the two-line chain. The relation between the normalised crack speed, $v/c$, and the internal energy level, $\Gg=\CE/\CE_c$ (the solid curves), found analytically based on \eq{gammmm} for $\Ga=0.2. 0.5, 1$. The small open circles situated on the vertical axis correspond to the static state with the semi-infinite bridge crack.
The other small open circles on the graphs correspond to the steady-state regimes found by numerical simulations of the corresponding transient problems.
The larger solid circle corresponds to the upper bound of the domain of the pure bridged crack. The larger open circle corresponds to the lower bound of the domain of the established partially bridged crack regime, where the length of the bridge zone depends on the level of the internal energy, $\Gg$, and the anisotropy parameter, $\Ga$. These two circles bound the intermediate regime, where the rate of the initially stretched bonds is uniform, while the breakage of the initially compressed bonds are chaotic.  }
\label{f4}
\end{figure}

\begin{figure}[h!]
  \begin{center}
    \includegraphics [scale=0.37]{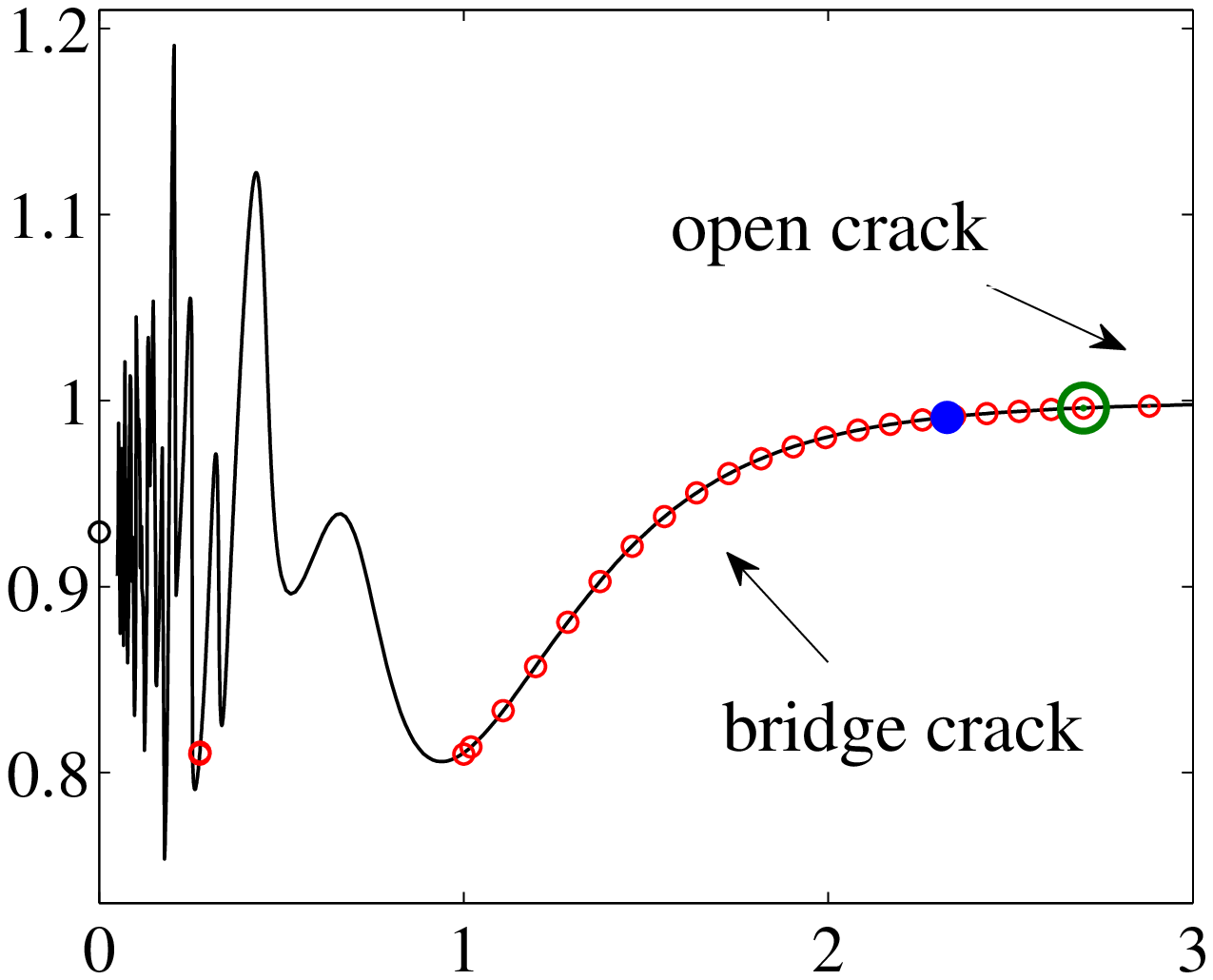}\includegraphics [scale=0.37]{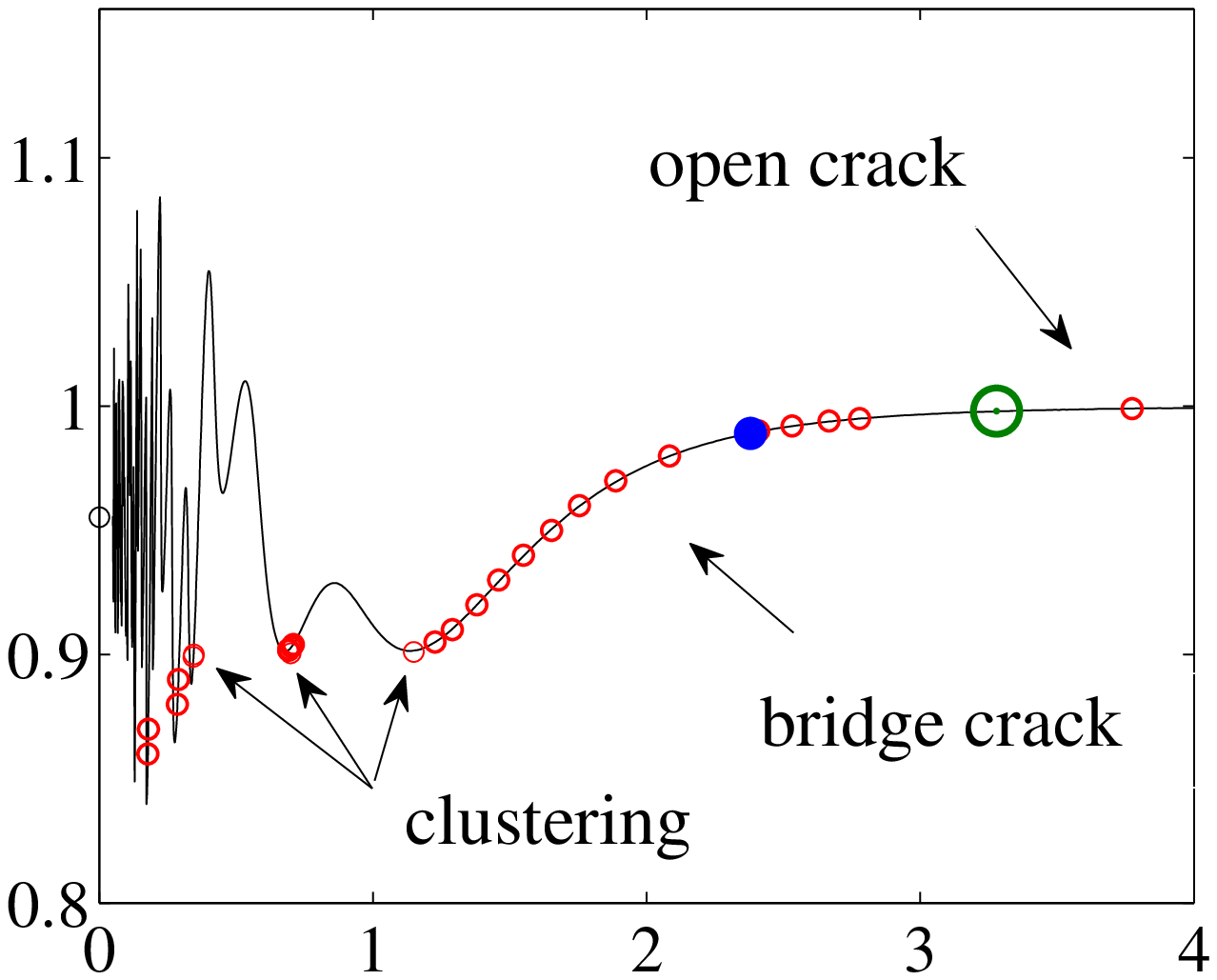}\includegraphics [scale=0.37]{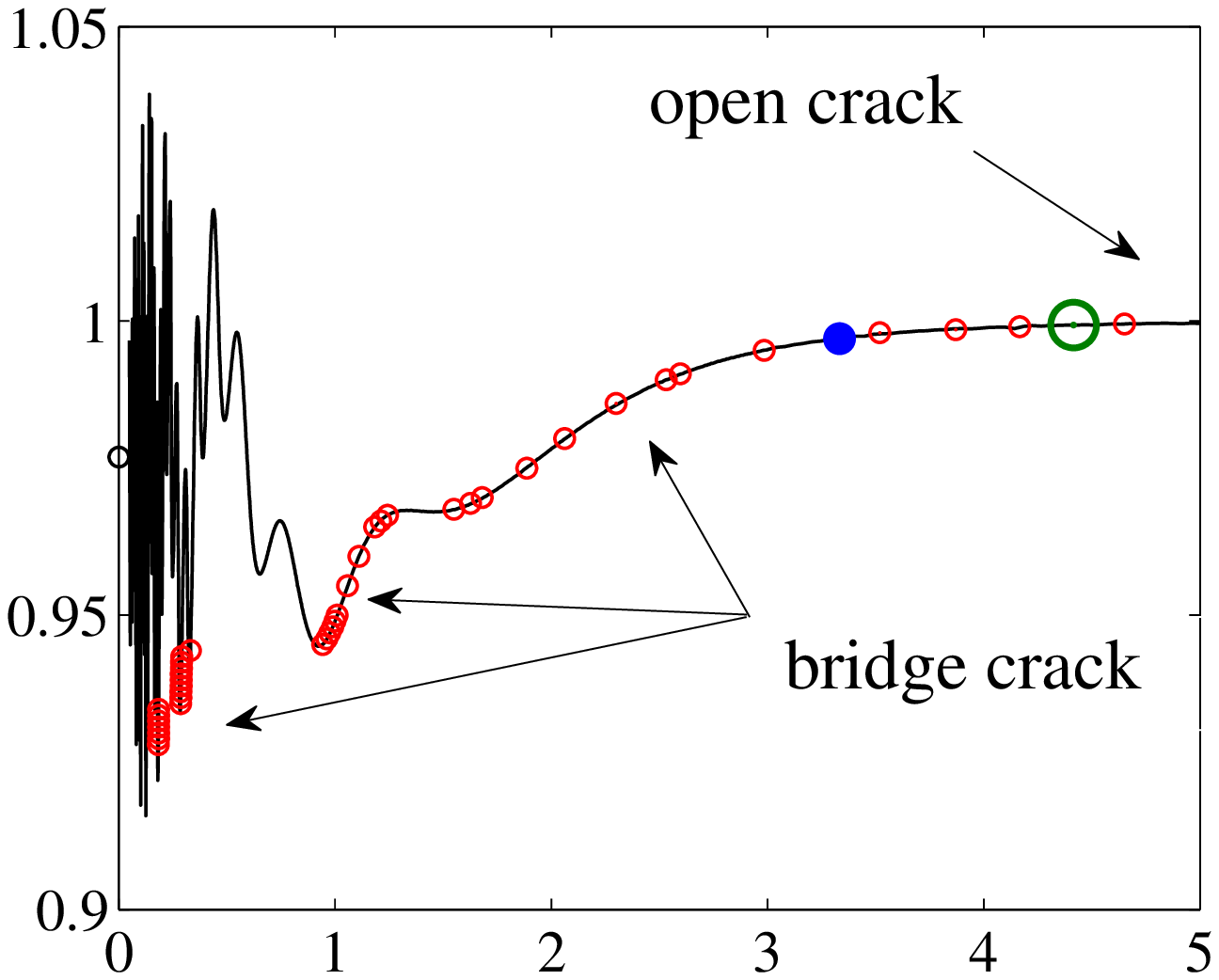}
       \put(-410,-5){\small$v/c$}\put(-250,-5){\small$v/c$}\put(-90,-5){\small$v/c$}
       \put(-480,80){$\gamma$}\put(-320,80){$\gamma$}\put(-160,80){$\gamma$}
        \put(-428,100){\small$\alpha=5.5$}\put(-285,100){\small$\alpha=9.6$}\put(-120,100){\small$\alpha=20$}
  \put(-480,120){a)}\put(-320,120){b)}\put(-160,120){c)}
  \end{center}
\caption{\small The $\Gg - v/c$ dependencies as in \fig{f4} for $\Ga=5.5, 9.6, 20$. Here low crack speed regimes with clustering are marked, where there exist crack speed oscillations within the cluster.}
\label{f5}
\end{figure}

\begin{figure}[h!]
  \begin{center}
    \includegraphics [scale=0.37]{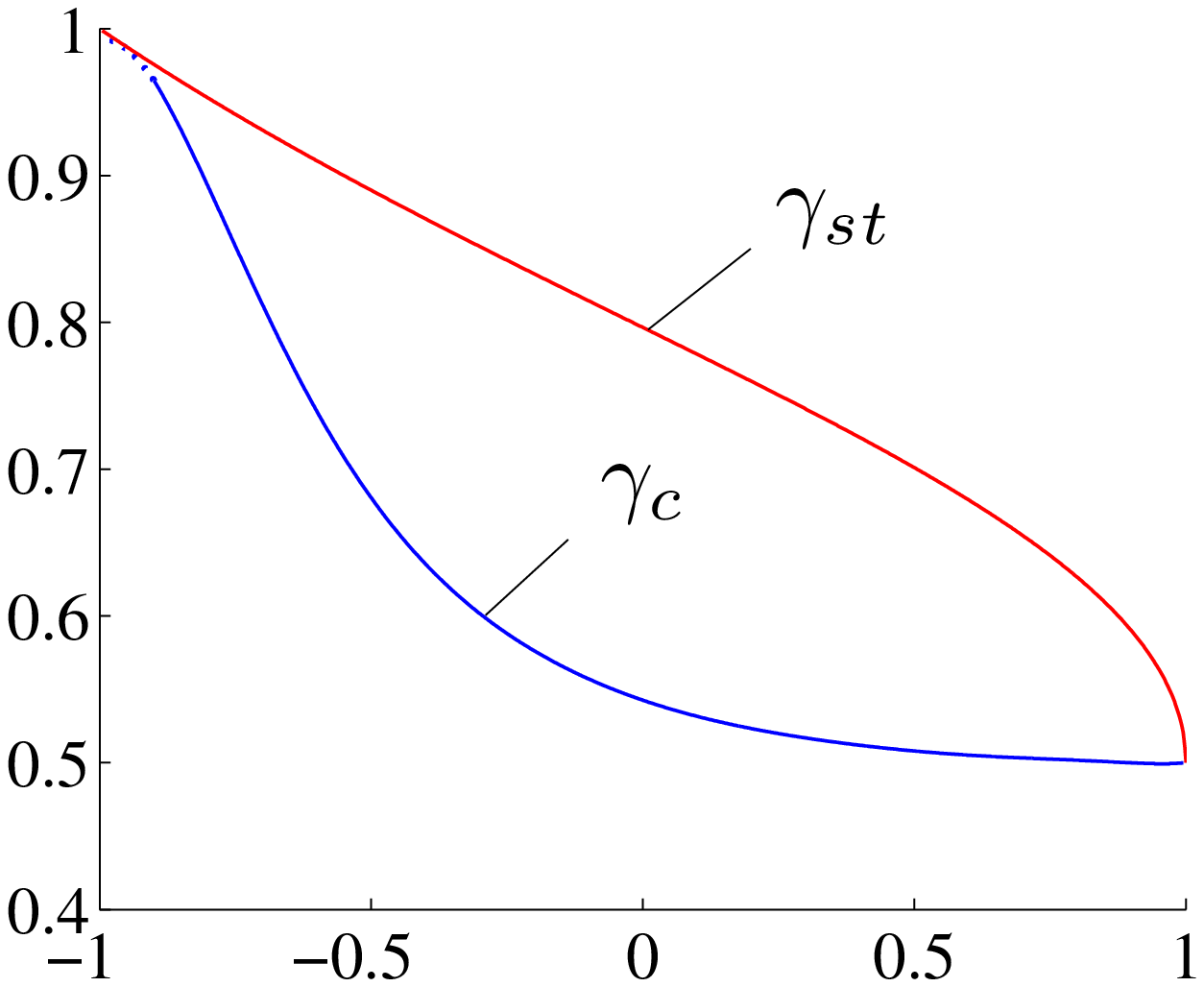}\includegraphics [scale=0.38]{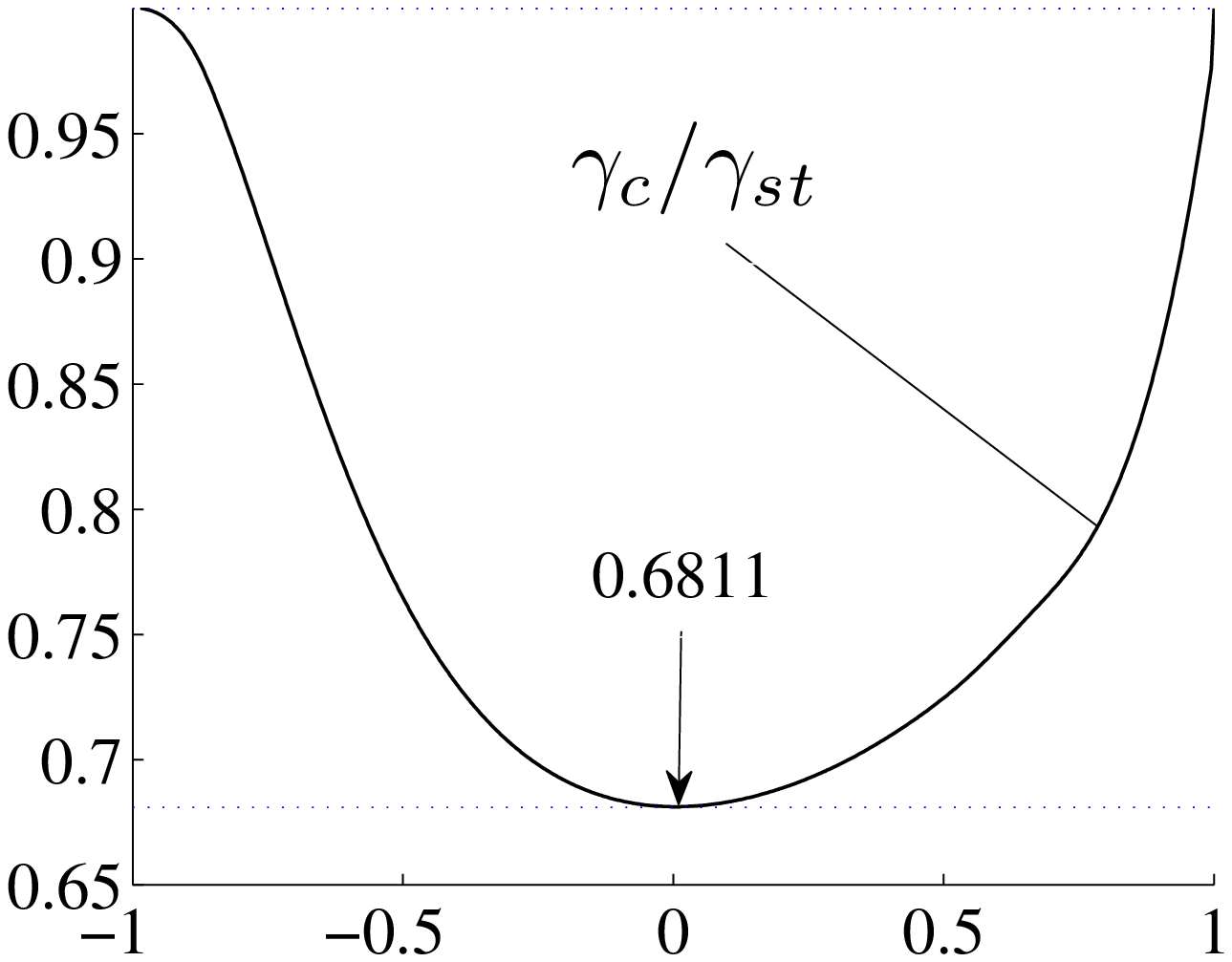}\includegraphics [scale=0.37]{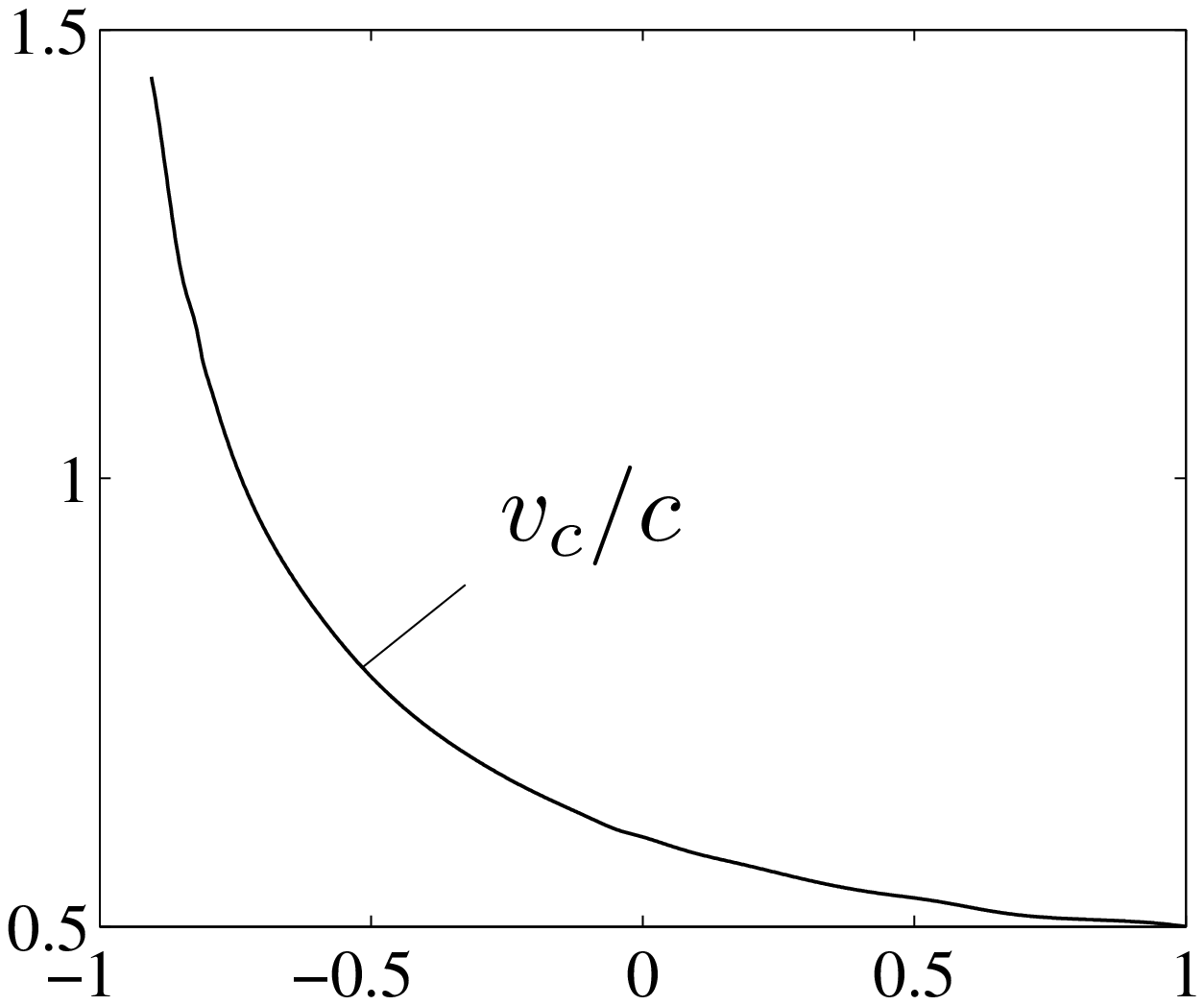}
      \put(-400,-5){\small$\hat\alpha$} \put(-240,-5){\small$\hat\alpha$}\put(-80,-5){\small$\hat\alpha$}
  \put(-480,120){a)}\put(-320,120){b)}\put(-160,120){c)}
  \end{center}
    \caption{\small The spontaneous semi-infinite bridged crack in the two-line chain. The comparative plots of the critical values of $\Gg_c(\hat{\Ga})$ (the minimal values of $\Gg$ under which the steady-state regime of the spontaneous crack growth exists) and the corresponding quasi-static dependence, $\Gg_{st}$ (a,b), and the crack speed, $v(\hat{\Ga})/c$, corresponding to $\Gg=\Gg_c$. Recall that $\Gg_c$ corresponds to the minimum $\Gg$ on the smooth parts of the plots in \fig{f4}. The manifestation of the dynamic amplification factor (see Slepyan (2000)), which leads to the considerable decrease of the minimal $\Gg$ in dynamics, can be clearly seen.}
\label{f6}
\end{figure}

The $\Gg - v/c$ relations for some values of $\Ga$ are presented in \fig{f4} $-$ \fig{f6}a. In these figures, (a) the results where $\Gg>1$ are not admissible, (b) the solid lines are plotted based on the analytical result  \eq{gammmm}, (c) the small circles situated on the vertical axis correspond to the static state with the semi-infinite bridge crack. In addition, the results of numerical simulations of the corresponding transient problem are reflected by the circles on the graphs.

Three different regimes were detected. The first one is the stable steady-state bridge crack growth as that predicted analytically. The upper bound of the corresponding speeds is marked by a larger bold circle. Next, there is a higher speed region, where the crack growth is accompanied by an irregular breakage of the initially compressed bonds; it is bounded by the larger open circle. Finally, at the right of the latter there exists the high-speed region, corresponding to the stable steady-state crack growth. In this region, however, only a finite bridge zone remains adjacent to the crack front, while the fully open crack front propagates with the same speed at a distance. Thus, the spontaneous crack growth regime changes as the internal energy approaches the critical value (that results in unlimited crack speed growth); however, the analytically and numerically obtained crack speeds remain equal for any energy level, for any crack speed. Below we discuss the findings of numerical simulations in more detail.

The lower bound of $\Gg$-region, $\Gg_c<\Gg<1$, where the spontaneous crack can propagate, as a function of the anisotropy parameter, $\hat{\Ga}$, versus the corresponding crack-initiation dependence, $\Gg_{st}$, obtained in Mishuris and Slepyan (2014) is presented in \fig{f6}a,b. The crack speed dependence corresponding to $\Gg=\Gg_c$ is shown in \fig{f6}c.

\begin{figure}[h!]

\vspace{5mm}

\begin{center}
 \includegraphics [scale=0.50]{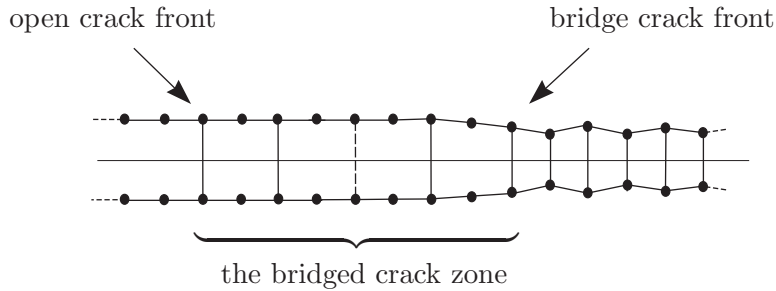}
\end{center}

\begin{picture}(0,60)(-550,-70)

    \put(-470,72){\small open crack front\hspace{43mm} bridge crack front}

\put(-400,-5){\small$\underbrace{\hspace{43mm}}$}
\put(-390,-25){\small the bridged crack zone}
\end{picture}

\vspace{-15mm}

    \caption{\small The chain structure in the steady-state open crack regime, where two fronts, the bridged crack front and the open crack front behind it, move with the same speed.}
\label{f_open}
\end{figure}

\section{The numerical simulations}\label{nss}
The spontaneous crack regimes were studied considering a sufficiently long two-line chain to prevent undesirable influence of their ends. For a given internal energy (given $\GD$) the initial state was taken in accordance with the static solution for the chain with a semi-infinite bridged crack found in Mishuris and Slepyan (2014). For the energy below critical in statics, $\Gg<\Gg_{st}$, where the state is stable, the first initially stretched bond was instantaneously removed, whereupon the particle movement was calculated with account taken of the further bond breakage under the critical strain, $u=u_c>0$. (Note that there is no need in the value of $u_c$ since the results presented in terms of the non-dimensional ratio, $\Gg=\CE/\CE_c$ are independent of it.) The numerical results are presented in \fig{f4} $-$ \fig{f6} and \fig{f112} $-$ \fig{f118}.

\subsection{Anisotropic chain corresponding to $\alpha=0.5$}
\begin{figure}[h!]
  \begin{center}
    \includegraphics [scale=0.69]{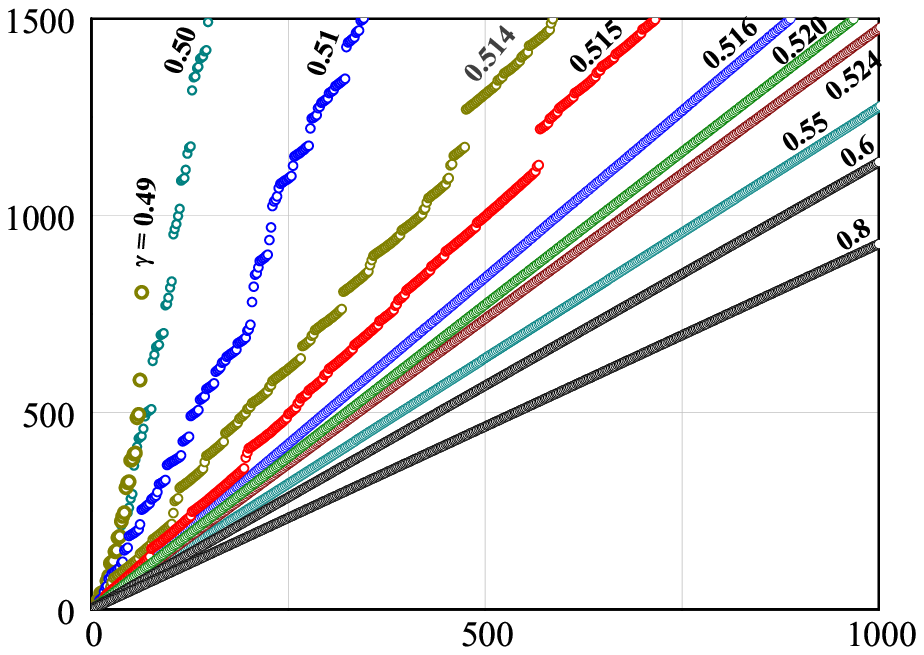}\hspace{20mm}\includegraphics [scale=0.57]{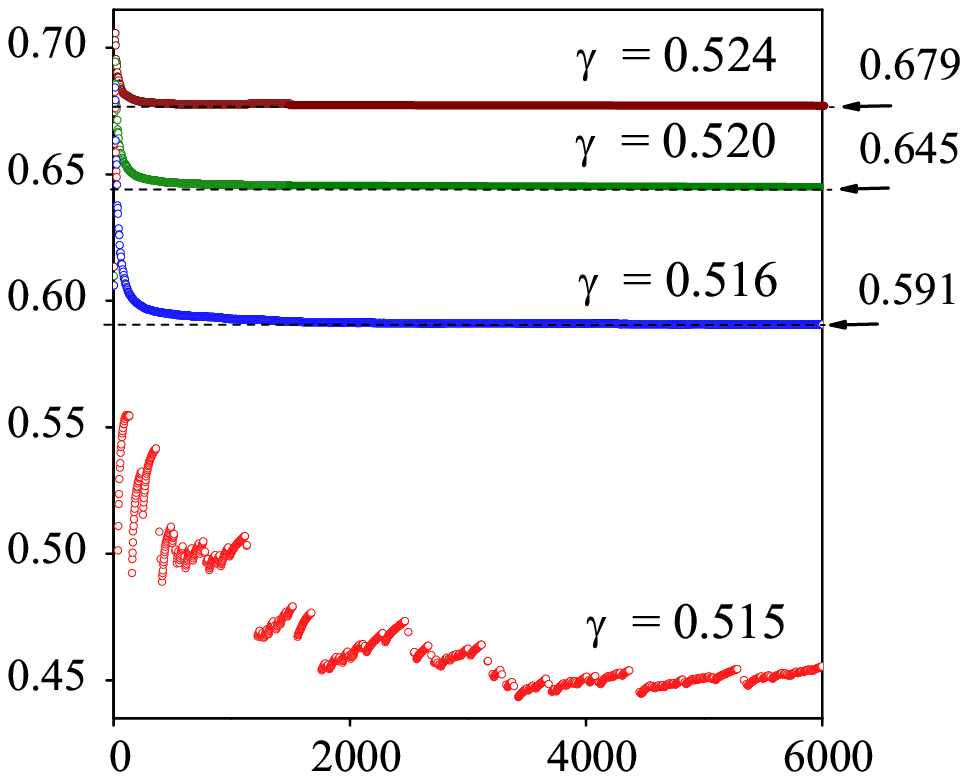}
       \put(-330,3){$m$}
       \put(-84,3){$t$}
       %\put(-480,80){$\gamma$}
       \put(-430,115){$t$}
       \put(-183,100){$\langle v\rangle$}
        \put(-420,150){a)}\put(-160,150){b)}
  \end{center}
    \caption{\small Results of the numerical simulations. The spontaneous bridge crack for $\Ga=0.5$. The plots for the current length of the pure bridged crack, $m(t)$ (a), and some of the corresponding average speeds (b). The lack-of-energy transient regimes for $0.500\le \Gg \le 0.515$ and the establishing steady-state modes for $\Gg\ge 0.516$.}
\label{f112}
\end{figure}

\begin{figure}[h!]
  \begin{center}
    \includegraphics[scale=1.4]{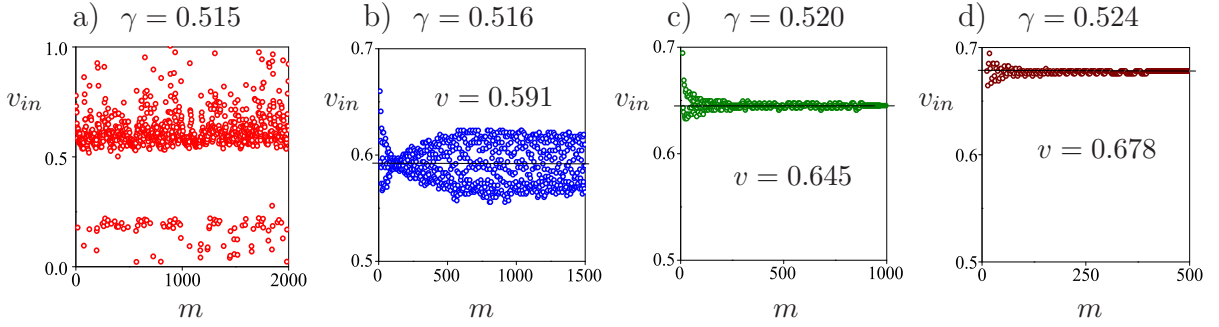}
    \begin{picture}(0,1)(-235,-160)
      \put(-390,0){\small$m$} \put(-280,0){\small$m$}\put(-165,0){\small$m$} \put(-50,0){\small$m$}
      \put(-455,80){\small$v_{in}$}\put(-335,80){\small$v_{in}$}\put(-225,80){\small$v_{in}$}\put(-110,80){\small$v_{in}$}
      \put(-293,80){\small$v=0.591$}

      \put(-180,50){\small$v=0.645$}
      \put(-65,60){\small$v=0.678$}
       \put(-430,110){a)~~\small$\gamma=0.515$}
       \put(-320,110){b)~~\small$\gamma=0.516$}
       \put(-205,110){c)~~~\small$\gamma=0.520$}
       \put(-95,110){d)~~~\small$\gamma=0.524$}
       \end{picture}

 \end{center}

      \vspace{-60mm}

    \caption{\small Results of the numerical simulations. The spontaneous bridge crack for $\Ga=0.5$. The current speed distributions. It is seen how the speed stabilises with growing internal energy and in time tending to the steady-state mode. }
\label{f113}
\end{figure}

The plots for the current length of the fully bridged crack as a function of time, $m(t)$-plots, and some of the corresponding average speeds are presented in \fig{f112} for a number of the internal energy levels. The steady-state regime exists for $\Gg\ge 0.516$, whereas the breakage in the range $0.50\le  \Gg\le 0.515$ continues but is not established. Note that the analytical minimum (see \fig{f4}b) is 0.515. The plots show that the steady-state regime is established as fast as the internal energy is large, and there is a wide gap between the averaged speeds of the steady-state, $\Gg>0.515$, and transient modes (see \fig{f112}b). The current speed distributions are shown in \fig{f113}. The high-amplitude crack speed oscillations about the averaged speed are seen existing at the minimal energy level and fast vanishing with the increase of the energy.

\vspace{5mm}

\subsection{Isotropic chain ($\alpha=1$)}
\begin{figure}[h!]
  \begin{center}
   \includegraphics[scale=1.8]{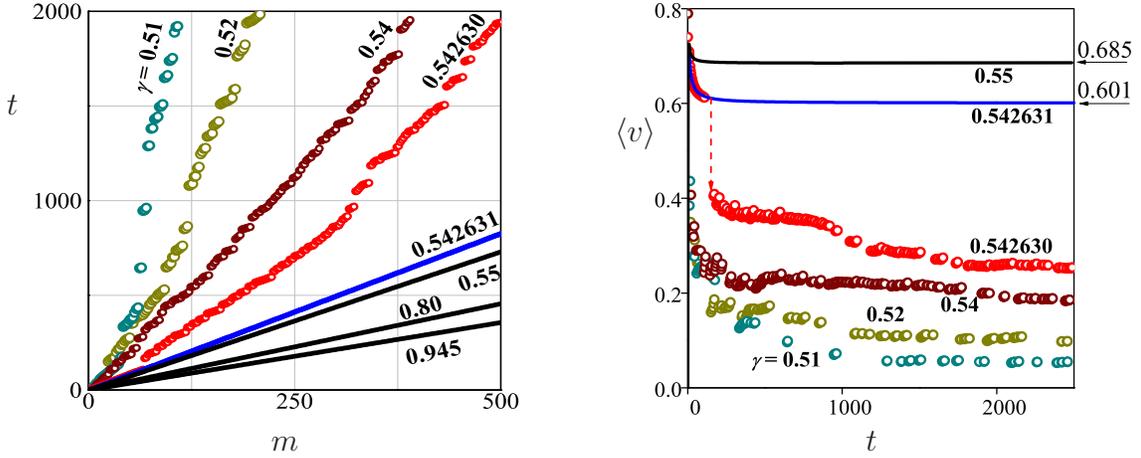}
  \begin{picture}(0,0)(0,-200)
       \put(-110,3){$m$}
       \put(115,3){$t$}
       \put(-210,130){$t$}
       \put(20,120){$\langle v\rangle$}
    \end{picture}
  \end{center}

      \vspace{-70mm}

    \caption{\small Results of the numerical simulations. The spontaneous bridge crack for $\Ga=1$. The plots for the current length of the fully bridged crack, $m(t)$ (a), and some of the corresponding average speeds (b). The extremely sharp interface can be observed between the steady-state and transient regimes at the energy minimum. It is seen how dramatically the speeds, both the averaged and current speeds, change under a barely noticeable increase in the energy level from $\Gg=0.542630$ to $0.542631$. In this tiny step, the crack propagation mode being almost chaotic becomes steady-state.}
\label{f115}
\end{figure}

\begin{figure}[h!]
  \begin{center}
    \includegraphics [scale=1.15]{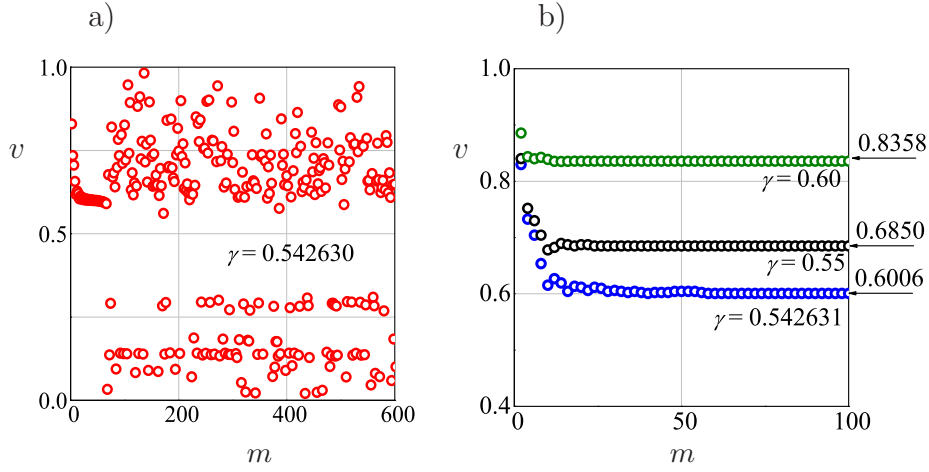}
      \begin{picture}(0,0)(50,-115)
      \put(-240,-25){\small$m$}\put(-80,-25){\small$m$}
      \put(-330,90){$v$}\put(-162,90){$v$}
       \put(-300,140){a)}\put(-130,140){b)}
       \end{picture}
  \end{center}
   \vspace{-30mm}

    \caption{\small Results of the numerical simulations. The spontaneous bridge crack for $\Ga=1$. The current speed distributions for some values of $\Gg$. The dramatic change is seen with a tiny step of the internal energy, $\Gg$ from $\Gg=0.542630$ to $0.542631$}
\label{f116}
\end{figure}

The plots of the crack tip position as a function of time and some of the corresponding average speeds are presented in \fig{f115} and \fig{f116} (compare with \fig{f112} and  \fig{f113} related to $\Ga=0.5$). The extremely sharp interface can be observed between the steady-state and transient regimes at the energy minimum. It is seen how dramatically the speeds, both the averaged and current speeds, change under a barely noticeable increase in the energy level from $\Gg=0.542630$ to $0.542631$. In this tiny step, the crack propagation mode being almost chaotic becomes steady-state. Note that the analytical minimum (see \fig{f4}c) is 0.5426.

\begin{figure}[h!]
  \begin{center}
    \hspace*{-15mm}\includegraphics [scale=2.4]{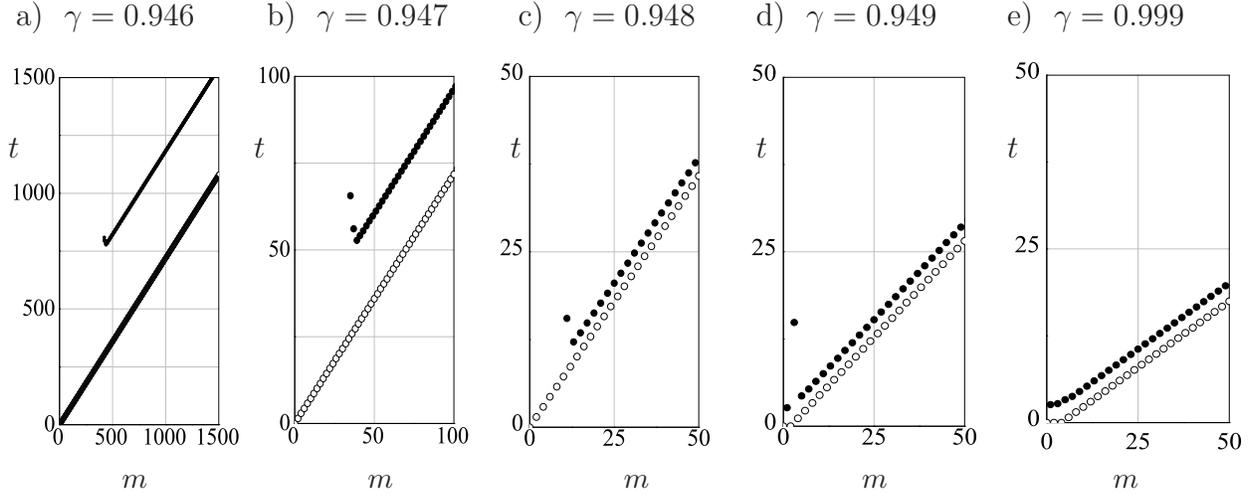}
    \begin{picture}(0,0)(-160,-290)
       \put(-350,30){\small$m$} \put(-257,30){\small$m$}\put(-160,30){\small$m$}\put(-70,30){\small$m$}\put(25,30){\small$m$}
       \put(-390,205){a)~~$\gamma=0.946$}
       \put(-295,205){b)~~$\gamma=0.947$}
       \put(-200,205){c)~~$\gamma=0.948$}
      \put(-110,205){d)~~$\gamma=0.949$} \put(-15,205){e)~~$\gamma=0.999$}

      \put(-393,155){$t$}\put(-300,155){$t$}\put(-203,155){$t$}\put(-110,155){$t$}\put(-15,155){$t$}

  \end{picture}
  \end{center}

  \vspace{-115mm}

    \caption{Results of the numerical simulations. The propagation of a partially bridged crack. The modes are presented where the initially compressed bonds break at an energy-dependent distance behind the front of the initially stretched bond breakage. In the numerically examined transient problem, the odd-bond breakage begins at a distance from the starting point of the dynamic crack. The latter distance as well as the former vanishes as the energy approaches the upper critical bound. In a general case of the steady-state regime, the length of the bridge crack zone is invariable. As can be seen in plots (c) there is a forward-backward splitting, the bonds break in the order $2m \to 2m-3 \to 2m+2\to 2m-1 \to ...$ .}
\label{f117}
\end{figure}

As the energy approaches the upper critical value, the compressed bonds begin to break at an energy-dependent distance behind the front. This is shown in \fig{f117}.  In the numerically examined transient problem, the odd-bond breakage begins at a distance from the starting point of the dynamic crack. The latter distance as well as the former vanishes as the energy approaches the upper critical bound. In a general case of the steady-state regime, the length of the bridge crack zone is time-independent, and its rear bound propagates with the same constant speed as the crack front. As can be seen in \fig{f117}c there is a forward-backward splitting, the bonds break in the order $2m \to 2m-3 \to 2m+2\to 2m-1 \to ...$ .

\subsection{Anisotropic chain corresponding to  $\alpha=9.6$}
\begin{figure}[h!]
  \begin{center}
    \includegraphics [scale=0.66]{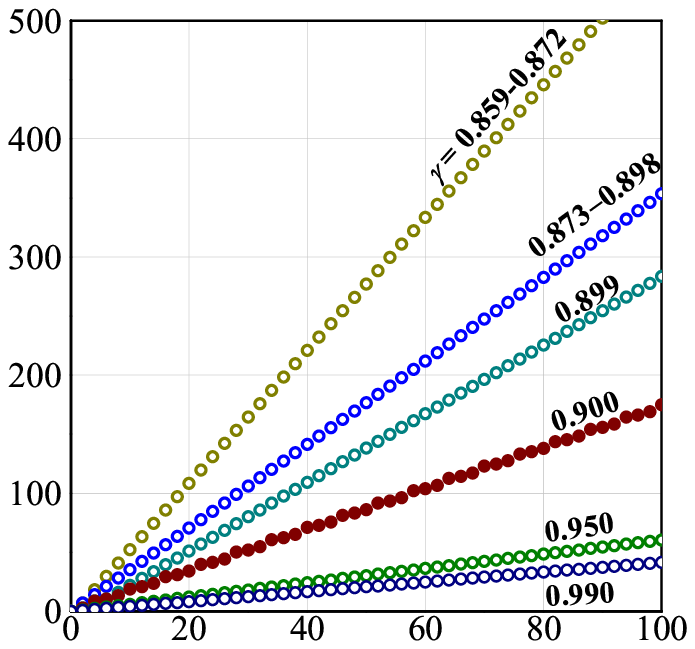}\includegraphics [scale=0.55]{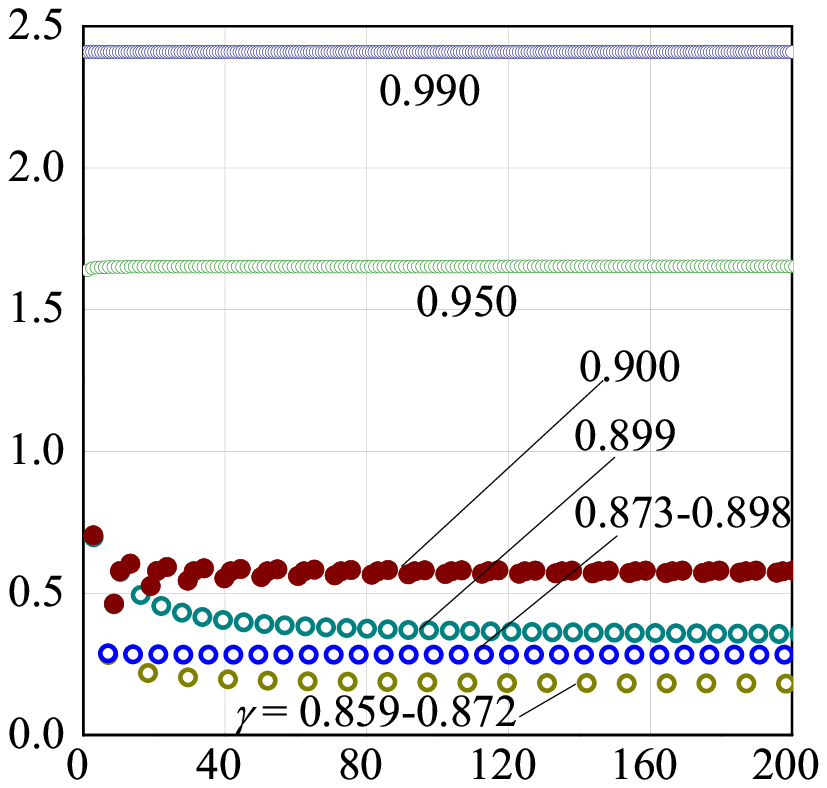}\includegraphics [scale=0.55]{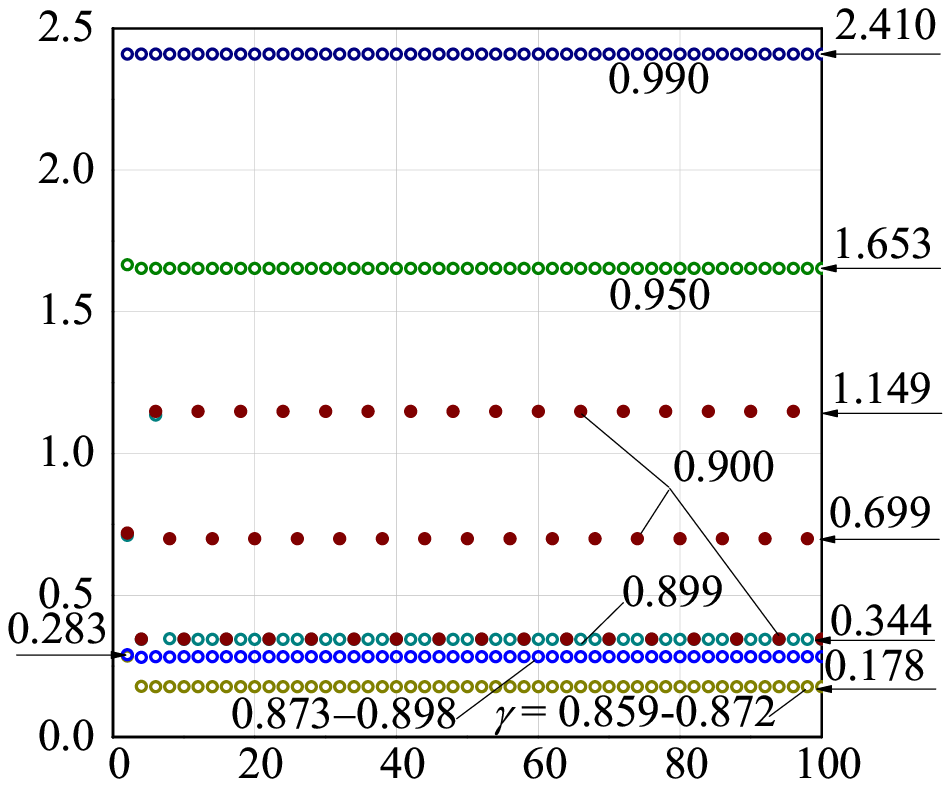}
       \put(-390,0){\small$m$}
       \put(-84,0){\small$t$}
       %\put(-480,80){$\gamma$}
       \put(-460,115){\small$t$}
       \put(-320,117){\small$\langle v\rangle$}
       \put(-160,117){\small$v$}
       \put(-240,0){\small$t$}

  \end{center}
    \caption{\small Results of the numerical simulations. The spontaneous bridge crack for $\Ga=9.6$. The plots for the current length of the fully bridged crack, $m(t)$ (a), and some of the corresponding average (b) and current (c) speeds. The three-even-bond clustering can be observed at $\Gg=0.900$. The current speeds in the cluster correspond to the minima shown in \fig{f5}b for the latter level of the energy. }
\label{f118}
\end{figure}

Such a stiff-transverse-bond structure presents us with some notable effects: (a) There exist here stable steady-state regimes corresponding to the oscillating part of the $v-\Gg$ diagram below the smooth minimum of the internal energy (compare \fig{f118} and \fig{f5}b). (b) In some of the energy ranges, the speed is energy-independent (see two upper lines in \fig{f118}a). This is because the regime corresponds to very narrow kinks in the $v-\Gg$ diagram as can be seen in \fig{f5}b. (c) The three-even-bond clustering arises at $\Gg=0.900$. The current speeds in the cluster correspond to the minima shown in \fig{f5}b for the latter level of the energy. The average and current speeds corresponding to the plots in \fig{f118}a are shown in \fig{f118}b and c, \res.

\subsection{Main results of the numerical simulations}
We now list the main findings of the numerical simulations.
\begin{itemize}
\item The spontaneous crack can propagate under a low level of the internal energy, even somewhat below the analytically found minimum; however, a weak, unstable regime corresponds to the latter case. It looks unsteady at the minimum but takes a stable, constant-speed mode with a small increase of the energy.
\item In the cases, where two or more values of the speed correspond to a smooth minimum of the energy as in \fig{f5},
the crack speed oscillates between the corresponding branches. So, in such case,  the crack violates the admissibility condition, which states that the only highest value of the speed is realised.
\item
The $v - \Gg$ diagram, $\Gg(v/c)$, at the left from its smooth minimum, $\Gg_c$, oscillates. As $\Ga$ grows, the oscillations become more intensive, and the global minimum of $\Gg$ appears below the $\Gg_c$, which is defined for the smooth part of the dependence. The numerical simulations show that there exists stable propagation regime corresponding to such sharp minima. In this regime,  the crack speed oscillates between two close branches as in the above-discussed case of a smooth minimum.
\item
For some $\Ga$ a double local minimum appears in the smooth part of the $v - \Gg$ dependence, which also results in the clustering with crack speed oscillations between the analytically found separate minima.
\item
At high levels of the internal energy, $\Gg$, the pure bridge-crack changes to partially bridged one, where only a finite bridge zone remains adjusted to the crack front, \fig{f_open}, while the fully open crack front propagates with the same speed at a distance. The length of the crack-bridge zone decreases to zero with the increase of $\Gg$. There also exists an intermediate region, where the bridged crack growth is accompanied by irregular breakage of the initially compressed bonds.
\item The regions of the internal energy corresponding to different crack speed regimes depend essentially on the anisotropy parameter $\alpha$.
\item The numerically found crack speeds coincide with those obtained analytically based on the steady-state bridged crack formulation. This also concerns the partially bridged and fully open crack regimes: the speed appears independent of the bridge zone length. Furthermore, the finite bridge zone bounds move with the same speed.
\item In the case of clustering with the crack speed oscillations, the instantaneous crack speeds in the cluster found numerically coincide with the corresponding values of the $v - \Gg$ analytical dependencies.
\end{itemize}
In the partially bridged crack regime, the right-directed crack velocity can be supersonic because no energy flux from the left is required, it takes the internal energy distributed along the crack path. At the same time, a disturbance, under which the energy barrier is overcome, goes from the breaking crack front to the neighboring bond at the right faster since the links between the masses are assumed massless. At the same time, under a supersonic crack speed any event, taking place at a several spans distance behind the crack front, cannot affect the crack speed. So, when the odd bonds break at a distance the even-bond breakage must propagate as predicted by the steady-state analytical solution obtained for the bridged crack, and this is confirmed by the numerical simulations.

\section{Conclusions}
Using the selective discrete transform introduced in Mishuris and Slepyan (2014) the analytical solution to the spontaneous steady-state crack propagation under the internal potential energy is derived. It corresponds to a general periodic structure, where the crack-path interface structure only is specified, and presented in terms of a unspecified dynamic Green's function. The latter is explicitly given for an anisotropic lattice and a two-line chain, and the problem for these structures are analytically considered in more detail.

The main results of the analytical solutions are the critical (minimal) energy levels, which admit the spontaneous steady-state crack, and the crack-speed $-$ energy relations as the $v/c - \Gg$ dependencies obtained for the lattice and chain under an arbitrary value of the anisotropy parameter, $\Ga$. The solutions evidence that the crack can propagate at subsonic as well as at supersonic speeds depending on the energy level. No upper bound of the speeds exists, whereas the lower bound does exist depending on $\Ga$.

For some values of $\Ga$ the analytical solutions also suggest crack-speed oscillation regimes at low energy levels, where the maximal-speed local minimum of the energy is not a single minimum or a global one.

Numerical simulations of the corresponding transient problem for the chain demonstrate the stability of the analytical solutions and validity of the $v/c - \Gg$ dependencies without exceptions. Also, the numerical simulations have revealed the existence of some more complicated regimes, such as (a) the disordered slow crack growth in an energy range below the lower bound corresponding to the steady-state formulation, (b) clustering with crack oscillations at the lower boundary, (c) breakage of the initially compressed bonds under a high level of the energy, and (d) independence of the crack speed of the breakage of the initially compressed bond.

Note that not only the spontaneous crack propagation under internal energy but the crack dynamics under both the internal energy and external forces is of interest. We consider the latter problem separately.

\vspace{5mm}
\noindent The authors {\bf acknowledge} support from the FP7 Marie Curie grant  No. 284544-PARM2.

\vskip 18pt
\begin{center}
{\bf  References}
\end{center}
\vskip 3pt

\inh Marder, M., and Gross, S., 1995. Origin of crack tip instabilities. J of the Mech. Phys. Solids 43, 1 - 48.

\inh Mishuris, G.S., Movchan, A.B., Slepyan, L.I., 2007. Waves and fracture in an inhomogeneous lattice structure. Waves in Random and
Complex Media, 17, 409-428. DOI: 10.1080/17455030701459910

\inh Mishuris, G.S., Movchan, A.B., Slepyan, L.I.,  2008. Dynamics of a bridged crack in a discrete lattice. The Quarterly Journal of
Mechanics and Applied Mathematics, 61, 151-160.

\inh Mishuris, G.S., Movchan, A.B., Slepyan, L.I., 2009. Localised knife waves in a structured interface. J. Mech. Phys. Solids 57, 1958-1979.

\inh
Mishuris, G.S., Movchan, A.B. and Bigoni, D. 2012. Dynamics of a fault steadily propagating within a structural interface.
SIAM Journal on Multiscale Modelling and Simulation, 10(3), 936–953.

\inh Mishuris, G.S., and Slepyan, L.I., 2014. Brittle fracture in a periodic structure with internal potential energy.  Proc. Roy. Soc. A  (in press).

\inh Munawar Chaudhri, M., 2009. The role of residual stress in a Prince Rupert's drop of soda-lime glass undergoing a self-sustained and stable destruction/fracture wave. Phys. Status Solidi A 206, No. 7, 1410–1413. DOI 10.1002/pssa.200925006

\inh Silverman, M.P., Strange, W, Bower, J., and Ikejimba, L., 2012. Fragmentation of explosively metastable glass. Phys. Scr. 85, 065403 (9pp). doi:10.1088/0031-8949/85/06/065403

\inh Slepyan, L.I., 1981. Dynamics of a crack in a lattice. Sov. Phys. Dokl., 26, 538-540.

\inh Slepyan, L.I., and Troyankina, L.V., 1984. Fracture Wave in a Chain Structure. J. Appl. Mech. Techn. Phys., 25, No 6, 921-927.

\inh Slepyan, L.I., 2000. Dynamic Factor in Impact, Phase Transition and Fracture. J. Mech. Phys. Solids, 48, 927-960.

\inh Slepyan, L.I., 2002. Models and Phenomena in Fracture Mechanics. Springer, Berlin.

\inh Slepyan, L.I., and Ayzenberg-Stepanenko, M.V., 2002. Some surprising phenomena in weak-bond fracture of a triangular lattice.
J. Mech. Phys. Solids 50(8), 1591-1625.

\inh Slepyan, L., Cherkaev, A. and Cherkaev, E, 2005.  Transition waves in bistable structures. II. Analytical solution: wave speed
and energy dissipation. J. Mech. Phys. Solids 53(2), 407-436.

\inh Slepyan, L.I., Mishuris, G.S., Movchan, A.B., 2010.  Crack in a lattice waveguide. Int J Fract 162, 91-106.
DOI: 10.1007/s10704-009-9389-5

\inh Vainchtein, A., 2010. The role of spinodal region in the kinetics of lattice phase transitions. J. Mech. Phys. Solids, 58(2): 227-240.

\end{document}